\documentclass[12pt]{article}

\usepackage{amsmath,amssymb,cite}
\usepackage{graphicx,color}

\def\bb#1{\mathbb{#1}}
\def\pare#1{\left( #1\right)}
\def\bpare#1{\left\{ #1\right\}}
\def\cpare#1{\left[ #1\right]}

\def\ds{\displaystyle}
\def\ssp{\hspace{0.3mm}}
\def\sn{\,\mathrm{sn}}
\def\cn{\,\mathrm{cn}}
\def\dn{\,\mathrm{dn}}
\def\iom{i\hspace{0.2mm}\omega}
\def\iomm#1{i\hspace{0.2mm}\omega_{#1}}
\def\tomm#1{i\hspace{0.2mm}\tilde\omega_{#1}}
\def\Re{\mathop{\rm Re}\nolimits\,}
\def\Im{\mathop{\rm Im}\nolimits\,}
\def\eK{\mathrm{\bf K}}
\def\eE{\mathrm{\bf E}}
\def\RS#1{{$\mathbb{R}\times S^{#1}$}}
\def\AdS#1{{$AdS_{#1}$}}
\def\AdSS#1{{$AdS_{#1} \times S^1$}}

\renewcommand{\eqref}[1]{$\pare{\rm \ref{#1}}$}

\newcommand{\f}[2]{\frac{#1}{#2}}
\newcommand{\ko}[1]{\left( #1 \right)}
\newcommand{\kko}[1]{\left[ #1 \right]}
\newcommand{\kkko}[1]{\left\{ #1 \right\}}
\newcommand{\abs}[1]{\left| #1 \right|}

\newcommand{\bmt}[1]{{{\mbox{\boldmath$ #1 $}}}}
\newcommand{\komoji}[1]{\mbox{$#1$}}

\def\cT{{\mathcal{T}}}
\def\cJ{{\mathcal{J}}}
\def\cE{{\mathcal{E}}}
\def\pa{\partial}
\def\eq{\equiv}

\def\sig{\sigma}
\def\al{\alpha}

\def\om{\omega}

\def\no{\nonumber}
\def\lam{\lambda}

\def\ep{{\epsilon}}
\def\ts{{\tau\leftrightarrow\sigma}}
\def\half{{\mbox{$\f{1}{2}$}}}

\newcommand{\CN}[2]{\,\mathrm{cn}\left( #1 , #2 \right)}

\DeclareMathOperator{\tr}{Tr}
\def\bul{{\small ~\,$\bullet$~\,}}
\numberwithin{equation}{section}

\begin{document}

\quad
\vspace{-2.0cm}

\begin{flushright}
{\bf September 2007}\\
\vspace{0.3cm}
{\small UT\,-\,07\,-\,26 \hfill \\
DAMTP\,-\,2007\,-\,83 \hfill \\}
\end{flushright}


\begin{center}
\Large\bf

Large Winding Sector of AdS/CFT

\end{center}

\renewcommand{\thefootnote}{$\alph{footnote}$}

\vspace*{0.5cm}
\centerline{
\sc
Hirotaka Hayashi$^{\dagger,\,}$\footnote{{\tt \,hirotakahayashi@hep-th.phys.s.u-tokyo.ac.jp}}\,,
\quad
Keisuke Okamura$^{\dagger,\,\ddagger,\,}$\footnote{{\tt \,okamura@hep-th.phys.s.u-tokyo.ac.jp}}\,,}
\vspace{3mm}
\centerline{
\sc
Ryo Suzuki$^{\dagger,\,}$\footnote{{\tt \,ryo@hep-th.phys.s.u-tokyo.ac.jp}}
\quad and\quad
Beno\^{i}t Vicedo$^{\ddagger,\,}$\footnote{{\tt \,B.Vicedo@damtp.cam.ac.uk}}}
\vspace*{0.5cm}
\begin{center}
${}^{\dagger}$\emph{Department of Physics, Faculty of Science,
University of Tokyo,\\
Bunkyo-ku, Tokyo 113-0033, Japan} \\
\vspace{0.3cm}
\vspace{0.3cm}
${}^{\ddagger}$\emph{DAMTP, Centre for Mathematical Sciences, Cambridge University,\\
Wilberforce Road, Cambridge CB3 OWA, UK} \\
\vspace*{0.5cm}
\end{center}

\renewcommand{\thefootnote}{\arabic{footnote}}
\setcounter{footnote}{0}

\centerline{\bf Abstract}

\vspace*{0.5cm}

We study a family of classical strings on $\mathbb R\times
S^{3}$ subspace of the $AdS_{5}\times S^{5}$ background that
interpolates between pulsating strings and single-spike strings.
They are obtained from the helical strings of {\tt hep-th/0609026}
by interchanging worldsheet time and space coordinates, which maps rotating/spinning string states with large
spins to oscillating states with large winding numbers. From a
finite-gap perspective, this transformation is realised as an
interchange of quasi-momentum and quasi-energy defined for the
algebraic curve. The gauge theory duals are also discussed, and
are identified with operators in the non-holomorphic sector of
$\mathcal N=4$ super Yang-Mills. They can be viewed as excited
states above the ``antiferromagnetic'' state, which is ``the farthest from BPS'' in the spin-chain spectrum. Furthermore, we
investigate helical strings on $AdS_{3}\times S^{1}$ in an
appendix.

\vspace*{1.0cm}

\vfill

\thispagestyle{empty}
\setcounter{page}{0}
\setcounter{footnote}{0}
\setcounter{figure}{0}
\renewcommand{\thefootnote}{\arabic{footnote}}
\newpage

\section{Introduction}

The AdS/CFT correspondence \cite{Maldacena:1997re} claims the type
IIB string theory on $AdS_{5}\times S^{5}$ is a dual description
of the four-dimensional, $\mathcal N=4$ super Yang-Mills (SYM)
theory. One of the predictions of the AdS/CFT is the exact
matching of the spectra on both sides, namely the conformal dimensions
of SYM operators with the energies of string states. In the
large\,-$N$ limit, these charges are supposed to be interpolated
by some function of the 't Hooft coupling $\lambda$\,, but the
strong/weak nature of the AdS/CFT usually prevents us from direct
comparison of the spectra.

Nevertheless, there has been considerable progress in matching the
spectra recently, based on the integrable structures of both theories.
They are captured by Bethe ansatz equations, which were was first
applied to the gauge theory side in the pioneering work of
\cite{Minahan:2002ve}. Despite the fact that we are lacking the
knowledge of perturbative computations for higher loop orders in
$\lam$ even for rather simple rank-one sectors, an all-order
asymptotic Bethe ansatz equation was proposed by assuming
all-order integrability as well as making use of some
sophisticated guesses \cite{Beisert:2004hm,Arutyunov:2004vx,Staudacher:2004tk,Beisert:2005fw}. There has been
increasing evidence and positive support for the conjectured
Bethe ansatz equation \cite{Beisert:2005tm,Janik:2006dc,Eden:2006rx,Arutyunov:2006iu,
Beisert:2006ib,Beisert:2006ez,Dorey:2007xn}, and significant
progress has been achieved in formulating the exact AdS/CFT Bethe
ansatz equation valid for all regions of $\lam$\,.

\paragraph{}
Many tests of the AdS/CFT conjecture in the large-$N$ limit have
taken place in the limit where a $U(1)_{\rm R}$\,-charge $J_{1}$
and conformal dimensions $\Delta$ of the SYM operators become very
large. The BMN limit \cite{Berenstein:2003gb} is one such
well-established limit. This limit is defined by sending $\lam$
to infinity while keeping $\lam'\eq \lam/J^{2}$ fixed, where
$J=J_{1}+{}$number of ``impurities''.

In \cite{Hofman:2006xt}, a different large-spin limit was
considered to serve as a new playground for the AdS/CFT. In this
limit, both $J_{1}$ and $\Delta$ go to infinity while the
difference $\Delta-J_{1}$ and the coupling $\lam$ are kept finite.
The worldsheet quantum corrections drop out in this limit, which
simplifies the comparison of both spectra considerably. Giant
magnons are string solutions living in this sector, which have an
infinite spin along one of great circles of $S^{5}$\,. They are
open objects, and the angular difference between the two endpoints on
the equator, which is equal to the localized worldsheet momentum,
is identified with the momentum of an excitation in the asymptotic
SYM spin-chain.

Giant magnons were generalized to the two-spin case in
\cite{Chen:2006ge} which carry an additional (finite) second spin
$J_{2}$\,, and are know as dyonic giant magnons. In static gauge,
the string equations of motion are essentially those of a bosonic
$O(4)$ sigma model supplemented by the Virasoro constraints, which
is classically equivalent to the Complex sine-Gordon (CsG) system.
Thus by using the Pohlmeyer-Lund-Regge (PLR) reduction procedure, the
dyonic giant magnon can be constructed as the counterpart of
a kink soliton solution of the CsG equation. In this connection, an
``elementary'' giant magnon of \cite{Hofman:2006xt} corresponds to
a kink soliton of the sine-Gordon (sG) equation. The SYM dual of the
dyonic giant magnon is a magnon boundstate in the asymptotic
spin-chain \cite{Dorey:2006dq,Chen:2006gp}, where the number of
constituent magnons corresponds to the second spin $J_{2}$ of the
string. It was shown that, in the large\,-$\lam$ limit, the
conjectured AdS/CFT S-matrix for boundstates  precisely agree with
the semiclassical S-matrix for scattering of dyonic giant magnons
under an appropriate choice of gauge \cite{Chen:2006gq}. For further
literature on giant magnons, see \cite{Spradlin:2006wk,GM,Okamura:2006zv,Ryang:2006yq} (See also \cite{Ryang:2005yd,Berenstein:2005jq,Vazquez:2006hd,Hatsuda:2006ty,Berenstein:2007zf}).
The idea of exploiting the relation between the classical CsG system
and the $O(4)$ string sigma model was further utilized to construct
more general classical strings, which are called helical strings
\cite{Okamura:2006zv}. They are the most general ``elliptic''
classical string solutions on $\mathbb R\times S^{3}$ that
interpolate between two-spin folded/circular strings
\cite{Frolov:2003xy} and dyonic giant magnons.

In the algebro-geometric approach to the string equations of
motion, these classical string solutions were studied as
finite-gap solutions. This line of approach stemmed from the work
\cite{Kazakov:2004qf}, and has provided many important
implications and applications in testing/formulating the
conjectured AdS/CFT S-matrix, including the quantum correction
\cite{Gromov:2007aq-Gromov:2007cd, Chen:2007vs}. In this
formalism, every string solution is characterized by a spectral
curve endowed with an Abelian integral called quasimomentum.
Recently helical strings were also reconstructed in this framework
\cite{Vicedo:2007rp} (see also \cite{Dorey:2006zj}). It enabled
us, in particular, to understand how folded/circular strings and
dyonic giant magnons interpolate from the standpoint of
algebraic curves.

\paragraph{}
In this paper, we investigate classical strings on an $\mathbb
R\times S^{3}$ subspace of $AdS_{5}\times S^{5}$ with large
winding numbers, rather than large spins. The recently found
single-spike solution of \cite{Ishizeki:2007we, Mosaffa:2007ty}
also falls into this category. In conformal gauge, they are obtained by performing a
transformation $\ts$ of large spin states,
{\em i.e.}, interchanging worldsheet time and space of
coordinates. Throughout this paper, we will refer to this
transformation as the ``$\ts$ transformation'', or just ``2D
transformation''. This kind of ``2D duality'' is well-known in the
context of rotating strings and pulsating string solutions, both of
which are characterized by the same special Neumann-Rosochatius
integrable system \cite{Arutyunov:2003uj,Arutyunov:2003za}. For example, if we
write the embedding coordinates of $S^{3} \subset \bb{R}^4$ as
$\xi_{j} = r_{j}(\tau,\sig) \; e^{i \varphi_j(\tau,\sig)}$
$(j=1,2)$ with sigma model constraint
$\sum_{j=1}^{2}\abs{\xi_{j}}^{2}=1$\,, the rotating strings are
obtained from the ansatz $r_{j}=r_{j}(\sig)$ and
$\varphi_{j}=w_{j}\tau + \al_{j}(\sig)$ with $w_{j}$ playing the
role of angular velocities, while pulsating strings follow from
the ansatz $r_{j}=r_{j}(\tau)$ and $\varphi_{j}=m_{j}\sig +
\al_{j}(\tau)$ with $m_{j}$ now representing the integer winding
numbers. It is reminiscent of T-duality that the angular momenta
(spins) and winding numbers are interchanged, however, one should
also take notice that not only the angular part $\varphi_{j}$ but
also the radial part $r_{j}$ are transformed in our case. To
summarize, there are two consequences of this $\ts$ map:
\begin{itemize}
\item Large spin states become large winding states.
\item Rotating/spinning states become oscillating states.
\end{itemize}
We will see these features for the case of 2D-transformed helical strings, and see how they interpolate between particular pulsating strings ($\ts$ transformed folded/circular strings) and the single-spike strings ($\ts$ transformed dyonic giant magnons).

It will be also shown that the two classes of string solutions \---- rotating/spinning with large-spins on the one hand, and oscillating strings with large windings on the other \---- correspond to two equivalence classes of representations of a generic algebraic curve with two cuts.
The $\ts$ operation turns out to correspond to rearranging the configuration of cuts with respect to two singular points on the real axis of the spectral parameter plane.\footnote{\,
An alternative description of $\ts$ operation is to swap the definition of quasi-momentum and so-called quasi-energy.
We will make this point clear later in Section \ref{sec:FG}.
}

\paragraph{}
Concerning the string/spin-chain correspondence of AdS/CFT, we
will claim that the dual operators of large-winding oscillating
strings are only found in a non-holomorphic sector. Such a
non-holomorphic sector has been much less explored than the
holomorphic, large-spin sectors, because of its intractability
mainly related with the non-closedness, or difficulty of
perturbative computations. Nevertheless, since our results,
together with the previous works
\cite{Okamura:2006zv,Vicedo:2007rp}, seem to complete the whole
catalog of classical, elliptic strings on $\mathbb R\times
S^{3}$\,, we hope they could shed more light not only on
holomorphic but also non-holomorphic sectors of the string/spin-chain duality,
for a deeper understanding of AdS/CFT. As a first step, in
Section \ref{sec:gauge theory}, we will identify the gauge theory
duals of the 2D transformed strings.

\paragraph{}
This paper is organized as follows. In Section \ref{sec:2D dual},
we briefly review the reduction of classical strings on $\mathbb
R\times S^{3}$ to the CsG system, and see the relation between helical
strings of \cite{Okamura:2006zv} and their 2D transformed version
from the CsG point of view. In Section \ref{sec:dual helical}, we
study 2D transformed versions of the type $(i)$ and type $(ii)$
strings. These new helical strings are interpreted as finite-gap
solutions in Section \ref{sec:FG}. In Section \ref{sec:gauge
theory}, we discuss the gauge theory interpretation of the 2D
transformed helical strings, and interpret them as excitations
above the ``antiferromagnetic'' state of the $SO(6)$ spin-chain.
Section \ref{sec:summary} is devoted to a summary and discussions.
In Appendix \ref{app:AdS helicals}, we present similar helical
solutions on $AdS_{3}\times S^{1}$\,. Some computational details
useful in discussing the infinite-winding limit can be found in
Appendix \ref{app:formula}.

\section{2D-transforming Classical Strings on \bmt{\mathbb R \times S^{3}}\label{sec:2D dual}}

We start with a brief review on how classical strings on $\mathbb R \times S^{3}$ are related to CsG system via the Pohlmeyer-Lund-Regge (PLR) reduction procedure \cite{PLR}, by summarizing the facts in \cite{Okamura:2006zv}.\footnote{\,
The notation used in this section basically follows from \cite{Okamura:2006zv}.
}
Then we see how the $\ts$ operation acts on the map.

Let us write the metric on $\mathbb R\times S^{3}$ as
\begin{equation}
ds^{2}_{\mathbb R\times S^3} = -d\eta_{0}^{2}+\abs{d\xi_{1}}^{2}+\abs{d\xi_{2}}^{2}\,.
\end{equation}
Here $\eta_{0}$ is the AdS time, and the complex coordinates $\xi_{j}$ $(j=1,2)$ are defined by the embedding coordinates $X_{M=1,\dots,4}$ of $S^{3} \subset \bb{R}^4$ as
\begin{equation}
\xi_{1} = X_{1}+i X_{2} = \cos \theta \; e^{i \varphi_1}  \qquad {\rm and} \qquad \xi_{2} = X_{3}+i X_{4} = \sin \theta \; e^{i \varphi_2} \,.
\label{embed_coords}
\end{equation}
We set the radius of $S^3$ to unity so that $\sum_{M=1}^{4} X_{M}^{2}=\sum_{j=1}^{2}\abs{\xi_{j}}^{2}=1$\,.
The Polyakov action for a string which stays at the center of the $AdS_{5}$ and rotating on $S^{3}$ takes the form,
\begin{equation}
S_{\mathbb R\times S^3} = - \f{\sqrt{\lambda}}{2}\int d\tau\int\f{d\sigma}{2\pi}\kkko{
\gamma^{ab}\kko{- \, \pa_{a}\eta_{\ssp 0} \, \pa_{b}\eta_{\ssp 0}+\pa_{a}\vec\xi\cdot\pa_{b}\vec\xi^{*}} + \Lambda(|\vec\xi|^{2}-1)
}\,,   \label{RtS3_action}
\end{equation}
where we used the AdS/CFT relation $\al' = 1/ \sqrt{\lam}$\,, and $\Lambda$ is a Lagrange multiplier.
We take the standard conformal gauge, $\gamma^{\tau \tau} = -1$\,, $\gamma^{\sigma \sigma} = 1$ and $\gamma^{\sigma \tau} = \gamma^{\tau \sigma} = 0$\,.
Denoting the energy-momentum tensor which follows from the action (\ref{RtS3_action}) as $\cT_{ab}$\,, the Virasoro constraints are imposed as
\begin{equation}\label{string_Virasoro}
\begin{matrix}
&\quad 0 &= \ds \cT_{\sig\sig}=\cT_{\tau\tau} = - \frac12 \, (\pa_{\tau}\eta_{0})^{2} - \frac12 \, (\pa_{\sigma}\eta_{0})^{2} + \frac12 \, |\pa_{\tau}\vec\xi|^{2} + \frac12 \, |\pa_{\sig}\vec\xi|^{2}\,, \\[3mm]
\mbox{and} &\quad 0&=\cT_{\tau\sig}=\cT_{\sig\tau}= \Re \pare{ \pa_{\tau}\vec\xi\cdot \pa_{\sig}\vec\xi^{*} }\,. \hfill
\end{matrix}
\end{equation}
The equations of motion that follow from \eqref{RtS3_action} are
\begin{equation}
\pa_{a}\pa^{a}\eta_{0}=0\quad
\mbox{and}\quad
\pa_{a}\pa^{\ssp a}\vec\xi+(\pa_{a}\vec\xi\cdot\pa^{\ssp a}\vec\xi^{*})\vec\xi = \vec 0\,.
\label{string_eom}
\end{equation}

\paragraph{}
It is well-known that the $O(4)$ (resp.\ $O(3)$) string sigma
model in conformal gauge is classically equivalent to Complex
sine-Gordon (resp.\ sine-Gordon) model with Virasoro constraints
\cite{PLR} (see also
\cite{Mikhailov:2005qv-Mikhailov:2005zd-Mikhailov:2005sy}). The
CsG system is defined by the Lagrangian
\begin{equation}
{\cal L}_{\rm CsG}=\frac{\partial_{a}\psi^{*}\, \partial^{a}\psi}{1-\psi^{*}\psi} + \psi^{*}\psi\,.
\label{CsG Lagrangian}
\end{equation}
The equation of motion (Complex sine-Gordon equation) which
follows from (\ref{CsG Lagrangian}) is
\begin{equation}
\partial_a \partial^{\ssp a} \psi  + \psi^{*} \, \frac{( \partial _a \psi )^2 }{1 - \psi^{*}\psi } - \psi \left( {1 - \psi^{*}\psi } \right) = 0\,.
\label{csg_eq}
\end{equation}
The PLR reduction relates the potential term $\partial_{a}\vec\xi\cdot\partial^{\ssp a}\vec\xi^{*}$ with a solution of CsG equation $\psi \equiv \sin \pare{\phi/2} \exp \pare{i \chi/2}$\,, as
\begin{equation}
\partial_{a}\vec\xi\cdot\partial^{\ssp a}\vec\xi^{*} = \cos \phi\,,
\label{PLR identify}
\end{equation}
and for each $\phi$\,, one can obtain a consistent classical string solution by solving a Schr\"{o}dinger type differential equation under appropriate boundary conditions.\footnote{\,
One can also trace back the PLR reduction procedure to obtain CsG solutions from classical string solutions.
}
For example, let us consider a kink soliton solution of CsG equation,
\begin{equation}
\psi (t,x) = \frac{\cos\al}{\cosh (x_v\cos\al )} \, \exp \pare{ i t_v \sin\al}\,,
\label{CsG kink}
\end{equation}
where $(t_v, x_v)$ are Lorentz-boosted coordinates
\begin{equation}
t_v \equiv \frac{t - v x}{\sqrt{1 - v^2}} \,,\qquad
x_v \equiv \frac{x - v t}{\sqrt{1 - v^2}} \,.
\label{def tvxv}
\end{equation}
Plugging the $\sin(\phi/2)$ part of \eqref{CsG kink} into \eqref{PLR identify}, and imposing the boundary condition
\begin{equation}
\xi_1 \to \exp \pare{\ssp i t \pm \Delta \varphi_1/2 \ssp},\quad \xi_2 \to 0, \qquad \pare{{\rm as}\ x \to \pm \infty},
\end{equation}
one reaches a dyonic giant magnon \cite{Chen:2006ge}.
In this case, the angular difference of two endpoints of the string $\Delta \varphi_1$ is determined through the CsG kink parameters $\al$ and $v$\,.

\paragraph{}
We are interested in how the 2D transformation acts on the dictionary.
Let us first look at the string equations of motion \eqref{string_eom} and the Virasoro constraints \eqref{string_Virasoro}. In view that they are invariant under the $\tau \leftrightarrow \sigma$ flip, any string solution is mapped to another solution under this map.
On closer inspection of the Virasoro constraints \eqref{string_Virasoro}, one actually finds that the $\ts$ operation can be applied independently to the $\bb{R} \subset AdS_5$ and $S^3 \subset S^5$ parts.
We will use this observation to generate new string solutions from known solutions on $\mathbb R\times S^{3}$\,, by transforming only the $S^{3}$ part while retaining the gauge $t \propto \tau$\,.
In order to satisfy other consistency conditions such as closedness of the string, one needs to care about the periodicity in the new $\sigma$ direction (that used to be the $\tau$ direction before the flip).

\paragraph{}
Before discussing the CsG counterparts of such $\ts$ transformed string solutions, it would be useful to review some relevant aspects of the (C)sG $\leftrightarrow$ string correspondence before the transformation.
A good starting point is the single-spin helical string constructed in \cite{Okamura:2006zv}.
It is a family of classical string on $\mathbb R \times S^{2}$ that interpolates between a folded/circular string of \cite{Gubser:2002tv} and a giant magnon.
From the standpoint of sG theory, the helical string corresponds to the following helical wave (``kink-train'') solution of sG equation,
\begin{equation}
\phi (t,x)=2\arcsin\kko{\CN{\f{(x-x_0) - v(t-t_0)}{k\sqrt{1-v^{2}}}}{k}}\,.
\label{helical cn_wave}
\end{equation}
via the PLR procedure. The single-spin helical string thus has two
controllable parameters derived from the sG soliton (\ref{helical
cn_wave})\,; one is the soliton velocity $v$ and the other is the
elliptic moduli parameter $k$ that controls the period of the
kink-array. In the $k\to 1$ limit, it reduces to an array of giant
magnons, while as $v\to 0$\,, it reduces to a folded/circular
string of \cite{Gubser:2002tv}.

\paragraph{}
Actually there is another periodic solution of sG equation, namely
a periodic instanton. Generally, one can interpret a static,
finite energy classical solution of sG theory in
$(1+1)$\,-dimensions as a finite action Euclidean solution in
$(1+0)$\,-dimension that interpolates between different vacua of
the theory. Such a sG instanton solution is known in the
literature (see, {\em e.g.}, \cite{PhysRevD.51.718}) and is given
by
\begin{equation}
\phi(t')=2\arcsin\kko{{\rm cn}\ko{\f{t'-t'_{0}}{k},k}}\,.
\label{period.inst.}
\end{equation}
Here $t'=it$ is the Euclidean time. One can see that a static kink
soliton of sG equation $-\pa_{x}^{2}\phi=\sin\phi$ (set $v=0$ in
(\ref{helical cn_wave})) is related to the instanton
(\ref{period.inst.}) of the Euclidean sG equation
$\pa_{it}^{2}\phi=-\pa_{t'}^{2}\phi=\sin\phi$ by a formal
translation $x\leftrightarrow t'$ ({\em i.e.}, space-like motion
turns into ``time-like'' motion), which  amounts to swapping
worldsheet variables $\tau\leftrightarrow\sig$\,. Starting from
the instanton solution (\ref{period.inst.})\,, and boosting it by
a parameter $v$\,, we obtain a one parameter family of sG
solutions of the form
\begin{equation}
\phi (t',x')=2\arcsin\kko{\CN{\f{(t'-t'_0) - v(x'-x'_0)}{k\sqrt{1-v^{2}}}}{k}}
\label{helical cn_wave-T}
\end{equation}
with $(t',x') = (it, ix)$\,,
which is related to the sG helical wave (\ref{helical cn_wave}) by
$\tau\leftrightarrow\sig$\,.

Via the PLR map, each periodic instanton corresponds to a
point-like segment, or ``string-bit'', and an infinite series of
such periodic sG instantons (\ref{period.inst.}) arrayed in the
$\sigma$-direction make up the corresponding classical string.
Note that for the boosted instanton (\ref{helical cn_wave-T}), $v$ no
longer represents a velocity, rather it should be viewed as a
parameter that controls the difference between time-origins
$t'_{0}$ for each bits. A pulsating string corresponds to the
$v=0$ case, when the timing of the pulsation of each string-bits
is perfectly right. When the pulsation timing of the bits is off
in a coherent manner, a symmetric ``spike'' comes into being,
reflecting the staggered motions of bits.\footnote{\, The
situation is much the same as the case of familiar transverse
waves, where oscillation in the medium takes place in a
perpendicular direction to its own motion. This direction of
motion corresponds to, in our case, the circumferential direction
along the equator of the sphere. } In the limit $k\to 1$\,, the
oscillation period of each bit becomes infinite, and the bits
stay in the vicinity of the equator for an infinite amount of
time, except during a short sudden jump away from the equator
\----\ this is one way to interpret the single-spin single-spike
string of \cite{Ishizeki:2007we} from the sG point of
view.\footnote{\, As is noticed in \cite{Ishizeki:2007we}, for sG
case, it is also possible to argue that the $\ts$ transformation
results in the change of sG kink soliton from $\phi = 2 \arcsin
\pare{1/\cosh x_v}$ to $\phi = 2 \arcsin \pare{\tanh x_v}$\,.
However, it seems this interpretation cannot be directly applied
to CsG case. }

\paragraph{}
We have just discussed the way to realise the oscillating
solutions resulting from a $\ts$ transformation in terms of a
collection of sG instantons. 
We gave this interpretation because it is very intuitive.
Actually one cannot generalise this
argument to the CsG case directly, since in this case the argument requires $\chi$ to be imaginary.
So for the CsG case, it would be convenient instead to interpret the
effect of the $\ts$ operation as flipping the sign of the ``mass''
term in the Lagrangian as
\begin{equation*}
{\cal L}_{\rm CsG}=\frac{\partial_{a}\psi^{*}\, \partial^{a}\psi}{1-\psi^{*}\psi} + \psi^{*}\psi
\quad \mapsto\quad \frac{\partial_{a}\psi^{*}\, \partial^{a}\psi}{1-\psi^{*}\psi} - \psi^{*}\psi\,.
\end{equation*}
In this way one can easily understand how one solution of CsG is related to another via the $\ts$ transformation (keeping $\phi$ and $\chi$ real).

Notice also, as in the soliton cases, that there are two classes of
``boosted'' instantons possible; the first is an instanton that
oscillates about one of the barriers of the periodic potential
with fixed finite oscillation range, while the other no longer
oscillates back and forth but goes on from one barrier to the
neighboring one. A similar kind of distinction exists for what we call type $(i)'$ and type $(ii)'$ strings.

\section{Helical Oscillating Strings\label{sec:dual helical}}

We are now in a position to discuss the 2D transformed helical
strings. We first study the type $(i)'$ case in the following
section \ref{sec:type (i)'}. The results on the type $(ii)'$
solutions will be collected in section \ref{sec:type (ii)'}.

\subsection{Type \bmt{(i)'} Helical Strings\label{sec:type (i)'}}

For the reader's convenience, let us display the profile of the
two-spin helical string obtained in
\cite{Okamura:2006zv},\footnote{\, Throughout this paper, we often
omit the elliptic moduli $k$ from expressions of elliptic
functions. For example, we will often write $\Theta_{\nu}(z)$ or
$\eK$ instead of $\Theta_{\nu}(z,k)$ or $\eK(k)$\,. }
\begin{align}
\eta_{0}^{\rm orig} &= a T + b X\,,   \label{zf0}  \\[2mm]
\xi_{1}^{\rm orig} &= C \frac{\Theta _0 (0)}{\sqrt{k} \, \Theta _0 (\iomm{1})} \frac{\Theta _1 (X   - \iomm{1})}{\Theta _0 (X )} \,
\exp \Big(  Z_0 (\iomm{1})  X+ i  u_1 T\Big)\,,
\label{zf1} \\[2mm]
\xi_{2}^{\rm orig} &= C \frac{\Theta _0 (0)}{\sqrt{k} \,\Theta _2 (\iomm{2})} \frac{\Theta _3 (X   - \iomm{2})}{\Theta _0 (X )} \, \exp \Big( Z_2 (\iomm{2} )X + i   u_2 T\Big)\,,
\label{zf2}
\end{align}
where $\omega_1$ and $\omega_2$ are real parameters, $k$ is the
elliptic modulus, and $C$ is the normalization constant given by
\begin{equation}
C = \pare{ \frac{\dn^2 (\iomm{2})}{k^2 \cn^2 (\iomm{2})} -
\sn^2 (\iomm{1}) }^{-1/2}\,.
\end{equation}
The coordinates $(T, X)$ are defined by
\begin{equation}
T = \frac{\tilde \tau - v \tilde \sigma}{\sqrt{1-v^2}} \,,\quad X = \frac{\tilde \sigma - v \tilde \tau}{\sqrt{1-v^2}} \,, \qquad (\tilde \tau, \tilde \sigma) \equiv (\mu\tau, \mu\sigma)
\label{boosted coordinates}
\end{equation}
with $\mu$ constant. Starting from (\ref{zf0})-(\ref{zf2}), by
swapping $\tau$ and $\sigma$ in $\xi_{i}(\tau, \sigma)$ $(i=1,2)$
while keeping the relation $\eta_{0}(\tau, \sigma)=aT+bX$ as it
is, one obtains the 2D-transformed version of the type $(i)$
two-spin helical strings, which we call type $(i)'$ helical
strings,
\begin{align}
\xi_{1} &= C \frac{\Theta _0 (0)}{\sqrt{k} \, \Theta _0 (\iomm{1})} \frac{\Theta _1 (T   - \iomm{1})}{\Theta _0 (T)} \,
\exp \Big(  Z_0 (\iomm{1})  T+ i  u_1 X\Big)\,,
\label{zf1-T} \\[2mm]
\xi_{2} &= C \frac{\Theta _0 (0)}{\sqrt{k} \,\Theta _2 (\iomm{2})} \frac{\Theta _3 (T   - \iomm{2})}{\Theta _0 (T)} \, \exp \Big( Z_2 (\iomm{2} )T + i   u_2 X\Big)\,.
\label{zf2-T}
\end{align}
The Virasoro constraints (\ref{string_Virasoro}) fix the parameters $a$ and $b$ in \eqref{zf0},
\begin{align}
a^2 + b^2 &= k^2 - 2 k^2 \sn^2 (\iomm{1}) - U + 2 \ssp u_2^2 \,,
\label{a,b-f1}  \\
\quad a \ssp b &= - i \, C ^2
\pare{u_1 \sn (\iomm{1}) \cn (\iomm{1}) \dn (\iomm{1}) - u_2 \, \frac{1-k^2}{k^2} \, \frac{\sn (\iomm{2}) \dn (\iomm{2})}{\cn^3 (\iomm{2})} }\label{a,b-f2}\,.
\end{align}
We can adjust the parameter $v$ such that the AdS time is proportional to the worldsheet time variable, namely $\eta_{0} = \sqrt{a^2 - b^2} \, \tilde\tau $ with $v \equiv b/a \le 1$\,.
The PLR reduction relation (\ref{PLR identify}) becomes
\begin{equation}
\frac{1}{\mu^2} \sum_{i=1}^2 \pare{ \left| \partial_\sigma \xi_{i} \right|^2
- \left| \partial_\tau \xi_{i} \right|^2 } = - k^2 + 2 k^2 \sn^2 (T) + U\,,
\end{equation}
which imposes the following constraints among the parameters
\begin{equation}
u_1^2 = U + \dn^2 (\iomm{1}) \,,\qquad
u_2^2 = U - \frac{(1 - k^2) \sn^2 (\iomm{2})}{\cn^2 (\iomm{2})} \,.
\label{u_1 and u_2 f}
\end{equation}

\paragraph{}
We are interested in closed string solutions, which means we need to consider the periodicity conditions.
The period in $\sigma$\,-direction is defined such that it leaves the theta functions in \eqref{zf1} and \eqref{zf2} invariant, namely it is given by
\begin{equation}
-\ell \le \sigma \le \ell,\qquad \ell = \frac{\eK \sqrt{1-v^2}}{v \ssp \mu} \,, \qquad \pare{v > 0}\,.
\end{equation}
Then, closedness of the string requires
\begin{alignat}{2}
\Delta \sigma &\equiv \frac{2 \pi}{n} & &= \frac{2 \eK \sqrt{1 - v^2}}{v \ssp \mu} \,,
\label{Dsigma_cl f} \\[2mm]
\Delta \varphi_1 &\equiv \frac{2 \pi N_1}{n} & &= 2 \eK \pare{ \frac{u_1}{v} + i Z_0 (\iomm{1}) } + (2 \ssp n'_1 + 1) \pi  \,,
\label{Dphi1_cl f} \\[2mm]
\Delta \varphi_2  &\equiv \frac{2 \pi N_2}{n} & &= 2 \eK \pare{ \frac{u_2}{v} + i Z_2 (\iomm{2}) } + 2 \ssp n'_2 \ssp \pi \,,
\label{Dphi2_cl f}
\end{alignat}
where $n=1,2,\dots$ counts the number of periods in $0\leq \sigma \leq 2\pi$\,, and $N_{1, 2}$ are the winding numbers in $\varphi_{1,2}$\,-directions respectively.
The integers $n'_{1,2}$ specify the ranges of $\om_{1,2}$ respectively.\footnote{\,
When $\om_{i}$ are shifted by $2\eK'$\,, the integers $n'_{i}$ change by one while $\xi_{i}$ and $\Delta \varphi_{i}$ are unchanged.}

The energy $E=(\sqrt{\lam}/\pi)\,{\cal E}$ and spins
$J_{i}=(\sqrt{\lam}/\pi)\,{\cal J}_{i}$ $(i=1,2)$ of the string
with $n$ periods are obtained from the usual definitions
\begin{equation}
{\cal E} = \int_{-n\ell}^{\ssp n\ell} d \sigma\, \pa_{\tau}\eta_{0}\,,\qquad
{\cal J}_i = \frac12 \int_{-n\ell}^{n\ell} d \sigma \; {\rm Im} \pare{\xi_i^* \partial_{\tau} \xi_i}\,,
\end{equation}
which yield in the present case,
\begin{align}
{\cal E} &= \frac{n a (1 - v^2)}{v} \, \eK = \frac{n (a^2 - b^2)}{b} \, \eK \,,
\label{t-f E} \\[2mm]
{\cal J}_1  &= \frac{n \ssp C^2 \, u_1}{k^2 }\cpare{ \eE - \left( {\dn^2 (\iomm{1}) + \frac{i \ssp k^2}{v \ssp u_1} \sn(\iomm{1}) \cn(\iomm{1}) \dn (\iomm{1}) } \right) \eK } \,,
\label{t-f J1} \\[2mm]
{\cal J}_2  &= \frac{n \ssp C^2 \, u_2}{k^2 } \cpare{ - \eE - (1-k^2) \left(
\frac{\sn^2 (\iomm{2})}{\cn^2 (\iomm{2})} - \frac{i}{v \ssp u_2} \frac{\sn(\iomm{2})\dn(\iomm{2})}{\cn^3(\iomm{2})} \right)\eK}\,.
\label{t-f J2}
\end{align}
It is meaningful to compare the above expressions with the ones for the original type $(i)$ helical strings of \cite{Okamura:2006zv},
\begin{align}
{\cal E}^{\rm orig} &= n \ssp a \pare{1 - v^2} \eK = \frac{n (a^2 - b^2)}{a} \, \eK \,, \\[2mm]
{\cal J}_1^{\rm orig}  &= \frac{n \ssp C^2 \, u_1}{k^2 } \cpare{ { - \eE + \left( {\dn^2 (\iomm{1}) + \frac{i \ssp v \ssp k^2}{u_1 } \, \sn(\iomm{1}) \cn(\iomm{1}) \dn (\iomm{1}) } \right)\eK} } \,, \\[2mm]
{\cal J}_2^{\rm orig}  &= \frac{n \ssp C^2 \, u_2}{k^2 } \cpare{ \eE + (1-k^2) \left(
\frac{\sn^2 (\iomm{2})}{\cn^2 (\iomm{2})} - \frac{i \ssp v}{u_2} \frac{\sn(\iomm{2})\dn(\iomm{2})}{\cn^3(\iomm{2})} \right)\eK}\,.
\end{align}
If we regard ${\cal E}$ and ${\cal J}_{i}$ as functions of $v =
b/a$, the global charges of the transformed solutions are related
to the original ones by ${\cal E} (a,b) = - {\cal E}^{\rm orig}
(b,a)$ and ${\cal J}_i (v) = - {\cal J}_i^{\rm orig} (-1/v)$\,.
Similar relations are also true for the winding numbers given in
\eqref{Dphi1_cl f} and \eqref{Dphi2_cl f}, $N_i (v) = - N_i^{\rm
orig} (-1/v)$ $(i = 1,2)$\,. They are just a consequence of the
symmetry $a\leftrightarrow b$ the Virasoro constraints possess.
For example, if $(a, b) = (a_0, b_0)$ solves \eqref{a,b-f1} and
\eqref{a,b-f2}, then $(a, b) = (b_0, a_0)$ gives another solution.

Notice that in the limit $v \to 0$ ($\om_{1,2}\to 0$)\,, all the
winding numbers in \eqref{Dsigma_cl f}-\eqref{Dphi2_cl f} become
divergent (and so ill-defined), due to the fact that the $\theta$
defined in \eqref{embed_coords} becomes independent of $\sigma$\,.
Therefore, in this limiting case, we may choose $\mu$ arbitrarily
without the need of solving \eqref{Dsigma_cl f}, provided that
$N_1$ and $N_2$ are both integers.

\begin{figure}[htbp]
\begin{center}
\vspace{0.5cm}
\includegraphics[scale=0.9]{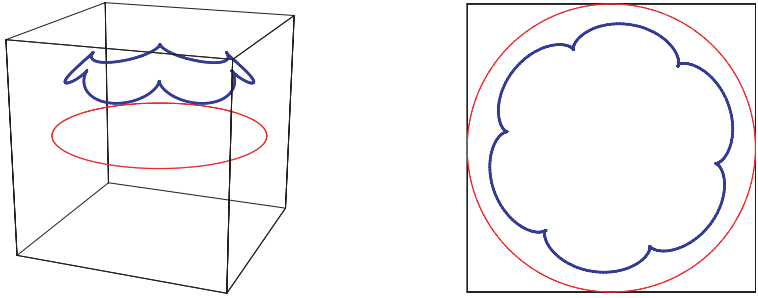}
\vspace{0.5cm} \caption{\small Type $(i)'$ helical string
$(k=0.68\,, n=6)$\,, projected onto $S^{2}$\,. The figure shows a
single-spin case $(u_{2}=\om_{2}=0)$\,. The (red) circle indicates
the $\theta=0$ line (referred to as the ``equator'' in the main
text).} \label{fig:T-I-gen}
\end{center}
\end{figure}

\paragraph{}
The type $(i)'$ helical strings contains both pulsating strings and single-spike strings in particular limits.
Below we will consider various limits including them.

\subsubsection*{$\bullet$ \bmt{\om_{1,2}\to 0} limit\,:~ Pulsating strings}

\begin{figure}[htbp]
\begin{center}
\vspace{0.5cm}
\includegraphics{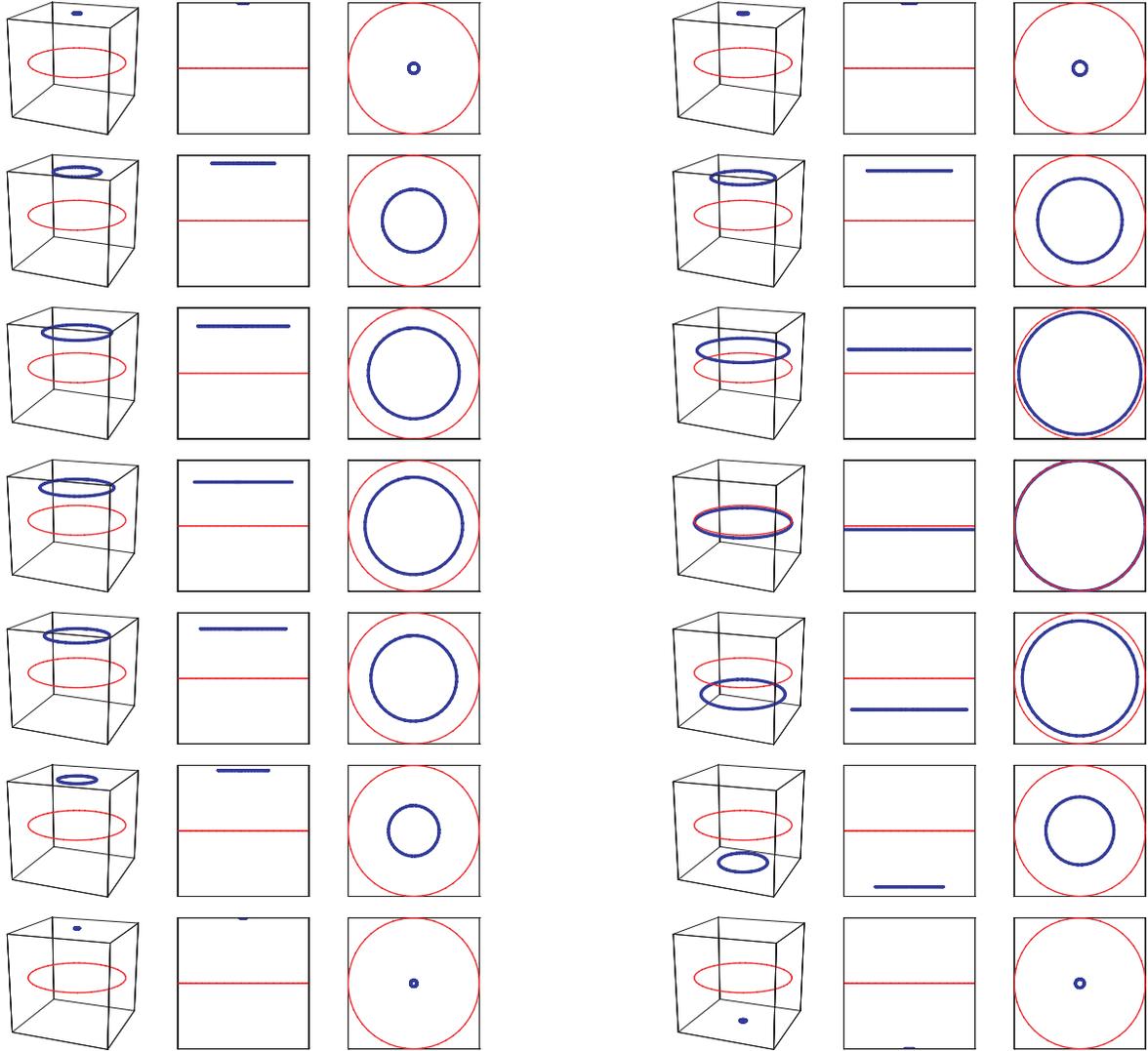}
\vspace{0.5cm} \caption{\small In the $\om_{1,2}\to 0$ limit, type
$(i)'$ (Left figure) and type $(ii)'$ (Right figure) helical
strings reduce to different types of pulsating strings. Their
behaviors are different in that the type $(i)'$ sweeps back and
forth only in the top hemisphere with turning latitude controlled
by the elliptic modulus, while the type $(ii)'$ pulsates on the
entire sphere, see Section \ref{sec:type (ii)'}. For the type
$(ii)'$ case, we only showed half of the oscillation period (for
the other half, it sweeps back from the south pole to the north
pole). } \label{fig:pulsating}
\end{center}
\end{figure}

Let us first consider the $\om_{1,2}\to 0$ limit. In this limit,
the boosted coordinates (\ref{boosted coordinates}) reduce to $(T,
X) \to (\tilde \tau, \tilde \sigma)$\,, and (\ref{zf0}),
(\ref{zf1-T})-(\ref{zf2-T}) become
\begin{equation}
\eta_{0} = \sqrt{k^2 + u_2^2} \ \tilde \tau \,, \qquad
\xi_{1} = k \sn(\tilde \tau, k) \, e^{i u_1 \tilde \sigma} \,, \qquad
\xi_{2} = \dn(\tilde \tau, k) \, e^{i u_2 \tilde \sigma} \,,
\label{profiles stat-f}
\end{equation}
with the constraint $u_{1}^{2}-u_{2}^{2}=1$\,. Since the radial
direction is independent of $\sigma$\,, we may treat $\mu$ as a
free parameter satisfying $N_1 = \mu u_1$ and $N_2 = \mu u_2$\,.
Then the conserved charges for a period become
\begin{align}
{\cal E} = \pi k \sqrt{N_1^2 + \pare{\frac{1}{k^2}-1} N_2^2} \,, \qquad
{\cal J}_1 = {\cal J}_2 = 0\,.
\label{charges stat-f}
\end{align}
Left of Figure \ref{fig:pulsating} shows the time evolution of the
type $(i)'$ pulsating string. It stays above the equator, and
sweeps back and forth between the pole $(\theta=\frac{\pi}{2})$
and the turning latitude determined by $k$\,.

When we set $u_{2}=0$\,, this string becomes identical to the
simplest pulsating solution studied in \cite{Minahan:2002rc} (the
zero-rotation limit of rotating and pulsating strings studied in
\cite{Engquist:2003rn, Kruczenski:2004cn}).\footnote{\, The type
$(i)'$ pulsating solution studied here and also the type $(ii)'$
pulsating string discussed later are qualitatively different
solutions from the so called ``rotating pulsating string''
\cite{Engquist:2003rn}, so that the finite-gap interpretation and
the gauge theory interpretation of type $(i)'$ and $(ii)'$ are
also different from those of \cite{Engquist:2003rn}. }

\subsubsection*{$\bullet$ \bmt{k\to 1} limit\,:~ Single-spike strings}

When the moduli parameter $k$ goes to unity, type $(i)'$ helical
string becomes an array of single-spike strings studied in
\cite{Ishizeki:2007we, Mosaffa:2007ty}. Dependence on $\omega_2$
drops out in this limit, so we write $\omega$ instead
$\omega_1$\,. The Virasoro constraints can be explicitly solved by
setting $a=u_{1}$ and $b=\tan\omega$\,. The profile of the string
then becomes
\begin{align}
\eta_{0} = \sqrt{1 + u_2^2}\, \tilde \tau  \,,\qquad
\xi_{1} = \frac{\sinh(T  - \iom)}{\cosh(T)} \, e^{i \tan(\omega)T + i u_1 X}\,,\qquad
\xi_{2} = \frac{\cos (\omega)}{\cosh(T)} \; \, e^{i u_2 X} \,.
\label{profiles spk}
\end{align}
with the constraint
$u_{1}^{2}-u_{2}^{2}=1+\tan^{2}\omega$\,.\footnote{\, Here
$u_{1,2}$ and $\om$ are related to $\gamma$ used in
\cite{Ishizeki:2007we} (see their Eq.\ (6.23)) by
$u_{1}=\f{1}{\cos\gamma\cos\om}$ and
$u_{2}=\f{\tan\gamma}{\cos\om}$\,. } The conserved charges are
computed as
\begin{equation}
{\cal E} = \pare{\frac{u_1^2 - \tan^2 \omega}{\tan \omega}} \eK (1) \,, \qquad
{\cal J}_1 = u_1 \cos^2 \omega \,, \qquad
{\cal J}_2 = u_2 \cos^2 \omega \,,
\label{charges spk}
\end{equation}
where $\eK (1)$ is a divergent constant.
For $n=1$ case (single spike), the expressions \eqref{charges spk} result in
\begin{equation}
{\cal J}_1 = \sqrt{ {\cal J}_2^2 + \cos^{2}\omega }\,,\qquad
{\it i.e.},\quad
J_{1}=\sqrt{J_{2}^{2}+\f{\lam}{\pi^{2}}\cos^{2}\omega}\,.
\label{spin spin}
\end{equation}

Since the winding number $\Delta \varphi_{1}$ also diverges as
$k\to 1$\,, this limit can be referred to as the ``infinite
winding'' limit,\footnote{\, Notice, however, that the string
wraps very close to the equator but touches it only once every
period (every ``cusp''). } which can be viewed as the
2D-transformed version of the infinite spin limit of
\cite{Hofman:2006xt}. By examining the periodicity condition
carefully, one finds that both of the divergences come from the
same factor $\left.\eK(k)\right|_{k\to 1}$\,. Using the formula
(\ref{leading Zeta}), one can deduce that
\begin{equation}
\left. {\mathcal E}-\f{\Delta\varphi_{1}}{2}\right|_{k\to 1}
=-\ko{\omega - \frac{\pare{2 \ssp n'_1 + 1}\pi}{2} } \equiv \bar \theta \,.
\label{ene-wind}
\end{equation}
Using the $\bar \theta$ variable introduced above, which is the same definition as used in \cite{Ishizeki:2007we}, one can see \eqref{spin spin} precisely reproduces the relation between spins obtained in \cite{Ishizeki:2007we}.

\begin{figure}[htbp]
\begin{center}
\vspace{0.5cm}
\includegraphics[scale=0.9]{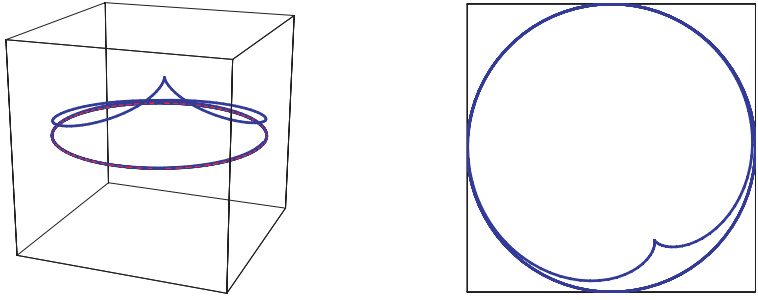}
\vspace{0.5cm}
\caption{\small The $k\to 1$ limit of type $(i)'$ helical string\,: single-spike string $(\omega=0.78)$\,.
The figure shows the single-spin case $(u_{2}=\om_{2}=0)$\,.}
\label{fig:T-I-ss}
\end{center}
\end{figure}

Let us comment on a subtly about $v \to 0$ (or equivalently $\om\to 0$) limit of a single spike string. It is easy to see the profile of single-spike solution \eqref{profiles spk} with $\omega=0$ agrees with that of pulsating string solution \eqref{profiles stat-f} with $k=1$\,, however, due to a singular nature of the $v \to 0$ limit, the angular momenta of both solutions \eqref{spin spin} and \eqref{charges stat-f} do not agree if we just naively take the limits on both sides.

\subsubsection*{$\bullet$ \bmt{k\to 0} limit\,:~ Rational circular (static) strings}

Another interesting limit is to send $k$ to zero, where elliptic functions reduce to rational functions.
The Virasoro conditions become
\begin{equation}
a^2 + b^2 = u_2^2 + \tanh^2 \omega_2 \quad
\mbox{and}\quad
ab = \pm u_2 \tanh \omega\,,
\label{uniform Vir}
\end{equation}
where $u_2 = \sqrt{U + \tanh^2 \omega}$\,.
This can be solved by $a = u_{2}$ and $b= \tanh \omega$ (assuming $U > 0$).
The profile is given by
\begin{equation}
\eta_{0} = \sqrt U \tilde \tau\,,\qquad
\xi_{1} =0\,,\qquad
\xi_{2} =e^{i \sqrt U \tilde \sigma}\,.
\label{rational static prof}
\end{equation}
This is an unstable string that has no spins and just wraps around one of the great circles, and can be viewed as the $\ts$ transformed version of a point-like, BPS string with $E-(J_{1}+J_{2})=0$\,.
The conserved charges for one period reduce to
\begin{equation}
\cE 
= \pi\mu \sqrt{U}\,, \qquad
\cJ_1=\cJ_2=0\,.
\label{(i)' k->0}
\end{equation}
The winding number for the $\varphi_{2}$\,-direction becomes
$N_{2}=\mu \sqrt{U}$\,, so the energy can also be written as 
\begin{equation}
E=N_{2}\sqrt{\lam}\,.
\label{winding energy}
\end{equation}
This result will be suggestive when we
discuss gauge theory later in Section \ref{sec:gauge theory},
since it predicts that the canonical dimension of SYM dual
operator, which should be the $SO(6)$ singlet state, is also given
by ${\rm (integer)}\times \sqrt{\lam}$ in this limit.
Note also that in the limit $\mu \sqrt{U} \to \infty$, the profile \eqref{rational static prof} agrees with the $\omega = \pi/2$ case of the single-spike string after the interchange $\xi_1 \leftrightarrow \xi_2$\,.
We will refer to this fact in the gauge theory discussion.

\subsubsection*{$\bullet$ \bmt{u_{2},\, \omega_{2} \to 0}\,: Single-spin limit}

A single-spin type $(i)'$ helical string is obtained by setting $u_{2}=\om_{2}=0$\,, which results in $J_2 = N_2 = 0$\,.\footnote{\,
It turns out the other single-spin limit $u_1\,,\omega_1 \to 0$\,, which gives $J_1 = 0$\,,  does not result in real solutions for this type $(i)'$ case.}
In view of \eqref{u_1 and u_2 f}, the condition $u_2 = \omega_2 = 0$ requires $U = 0$\,, $u_1 = \dn (\iom)$ and $C = \sqrt{k}/\dn (\iom)$\,, and the Virasoro constraints \eqref{a,b-f1} and \eqref{a,b-f2} are solved by setting $a = k \cn (\iom)$\,, $b = -ik \sn (\iom)$ and $v = - i \,\sn (\iom)/\cn (\iom)$\,.
Periodicity conditions then become
\begin{alignat}{3}
\Delta \sigma &= \ \frac{2 \pi}{n} & &= \frac{2 i \eK}{\mu \sn(\iom)} \,, \qquad
\frac{2 \pi N_2}{n} = 0\,, \\[2mm]
\Delta \varphi_1 &= \frac{2 \pi N_1}{n} & &= 2 i \eK \pare{ \frac{\cn(\iom) \dn (\iom)}{\sn (\iom)} + Z_0 (\iom) } + \pare{2 \ssp n'_1 + 1} \pi \,,
\label{closed single phi-i}
\end{alignat}
and the conserved charges for one period are
\begin{equation}
{\cal E} = \frac{i k}{\sn(\iom)} \, \eK \,,\qquad
{\cal J}_1 = \frac{1}{k \dn (\iom)} \Big[ \eE - \pare{1 - k^2} \eK \Big] \,,\qquad
{\cal J}_2 = 0\,.
\label{single ch-i}
\end{equation}

\subsection{Type \bmt{(ii)'} Helical Strings\label{sec:type (ii)'}}

The type $(ii)'$ solution can be obtained from the type $(i)'$ solutions, either by shifting $\omega_2\mapsto \omega_2+\eK'$ or by transforming $k$ to $1/k$\,.
The profile is given by\footnote{\,
We use a hat to indicate type $(ii)'$ variables.}
\begin{align}
\hat \eta_{0} &= \hat a \ssp T + \hat b X\,,   \label{zc0}  \\[2mm]
\hat \xi_{1} &= \hat C \frac{\Theta _0 (0)}{\sqrt{k} \, \Theta _0 (\iomm{1})} \frac{\Theta _1 (T - \iomm{1})}{\Theta _0 (T)} \, \exp \Big( Z_0 (\iomm{1}) T + i u_1 X\Big)\,,
\label{zc1} \\[2mm]
\hat \xi_{2} &= \hat C \frac{\Theta _0 (0)}{\sqrt{k} \,\Theta _3 (\iomm{2})} \frac{\Theta _2 (T - \iomm{2})}{\Theta _0 (T)} \, \exp \Big( Z_3 (\iomm{2} ) T + i u_2 X \Big)\,,
\label{zc2}
\end{align}
where $\hat C$ is the normalization constant,
\begin{equation}
\hat C = \pare{ \frac{\cn^2 (\iomm{2})}{\dn^2 (\iomm{2})} - \sn^2 (\iomm{1}) }^{-1/2}\,.
\end{equation}
The equations of motion force $u_{1}$ and $u_{2}$ to satisfy
\begin{equation}
u_1^2 = U + \dn^2 (\iomm{1}) \,,\qquad
u_2^2 = U + \frac{1 - k^2}{\dn^2 (\iomm{2})} \,,
\end{equation}
and the Virasoro conditions impose the following constraints between parameters $\hat a$ and $\hat b$\,,
\begin{alignat}{2}
&\hat a^2 + \hat b^2 &\ &= \quad k^2 - 2 k^2 \sn^2 (\iomm{1}) - U + 2 \ssp u_2^2 \,,
\label{a,b-c1}\\
&\quad \hat a \, \hat b &\ &= - i \, \hat C^2
\pare{u_1 \sn (\iomm{1})\cn(\iomm{1}) \dn(\iomm{1}) + u_2 \pare{1-k^2} \frac{\sn (\iomm{2})\cn(\iomm{2})}{\dn^3(\iomm{2})}} \,.
\label{a,b-c2}
\end{alignat}
As in the type $(i)'$ case, we can set $\hat \eta_{0} = \sqrt{\hat a^2 - \hat b^2}\,\tilde \tau$ with $\hat v \equiv \hat b/\ssp \hat a \le 1$\,.
The periodicity conditions for the type $(ii)'$ solutions become
\begin{alignat}{2}
\Delta \sigma &\equiv \frac{2 \pi}{m} = \frac{2 \eK \sqrt{1 - \hat v^2}}{\hat v \ssp \mu} \,,
\label{Dsigma2_cl_dy} \\[2mm]
\Delta \varphi_1 &\equiv \frac{2 \pi M_1}{m} = 2 \eK \pare{ \frac{u_1}{\hat v} + i Z_0 (\iomm{1}) } + \pare{2 \ssp m'_1 + 1} \pi  \,,
\label{Dphi3_cl_dy} \\[2mm]
\Delta \varphi_2 &\equiv \frac{2\pi M_2}{m} = 2 \eK \pare{ \frac{u_2}{\hat v} + i Z_3 (\iomm{2}) } + \pare{2 \ssp m'_2 + 1} \pi  \,,
\label{Dphi4_cl_dy}
\end{alignat}
where $m=1,2,\dots$ counts the number of periods in $0\leq \sigma
\leq 2\pi$\,, and $M_{1, 2}$ are the winding numbers in the
$\varphi_{1,2}$-directions respectively, and $m'_{1,2}$ are
integers. The conserved charges are given by
\begin{align}
\hat {\cal E} &= \frac{m a (1 - v^2)}{v} \, \eK = \frac{n (a^2 - b^2)}{b} \, \eK \,,
\label{t-c E} \\[2mm]
\hat {\cal J}_1  &= \frac{m \ssp \hat C^2 \, u_1 }{k^2} \cpare{ \eE - \left( {\dn^2 (\iomm{1}) + \frac{i \ssp k^2}{\hat v \ssp u_1} \, \sn (\iomm{1}) \cn (\iomm{1}) \dn (\iomm{1}) } \right) \eK } \,,
\label{t-c J1} \\[2mm]
\hat {\cal J}_2  &= \frac{m \ssp \hat C^2 \, u_2 }{k^2} \cpare{ - \eE + (1-k^2) \pare{ \frac{1}{\dn^2 (\iomm{2})} - \frac{i \ssp k^2}{\hat v \ssp u_2} \, \frac{\sn (\iomm{2})\cn(\iomm{2})}{\dn^3(\iomm{2})}} \eK }\,.
\label{t-c J2}
\end{align}

Just as in the type $(i)\leftrightarrow (i)'$ case, the winding
numbers and the conserved charges of the original type $(ii)$ and
$(ii)'$ are related by ${\cal \hat E} (\hat a, \hat b) = - {\cal \hat
E}^{\rm orig} (\hat b, \hat a)$\,, ${\cal \hat J}_i (\hat v) = - {\cal
\hat J}_i^{\rm orig} (-1/\hat v)$ and $M_i (\hat v) = - M_i^{\rm
orig} (-1/\hat v)$\,.

\begin{figure}[htbp]
\begin{center}
\vspace{0.5cm}
\includegraphics[scale=0.9]{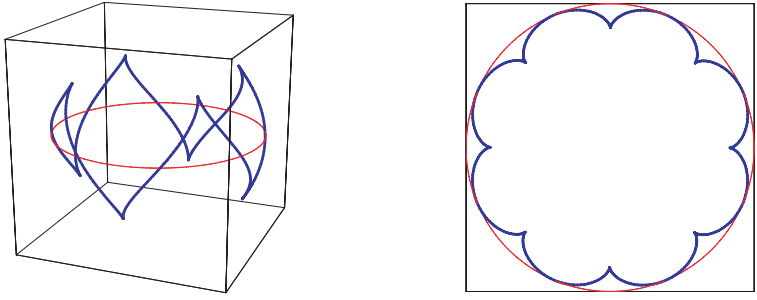}
\vspace{0.5cm}
\caption{\small Type $(ii)'$ helical string $(k=0.40\,, m=8)$\,.
The figure shows a single-spin case $(u_{2}=\om_{2}=0)$\,.}
\label{fig:T-II-gen}
\end{center}
\end{figure}

\paragraph{}
As in the type $(i)'$ case, we can take various limits.

\subsubsection*{$\bullet$ \bmt{\om_{1,2}\to 0} limit\,:~ Pulsating strings}

The profiles \eqref{zc0}-\eqref{zc2} reduce to
\begin{equation}
\hat \eta_{0} = \sqrt{1 + u_2^2} \ \tilde \tau \,, \qquad
\hat \xi_{1} = \sn(\tilde \tau, k) \, e^{i u_1 \tilde \sigma} \,, \qquad
\hat \xi_{2} = \cn(\tilde \tau, k) \, e^{i u_2 \tilde \sigma} \,,
\end{equation}
with constraint $u_{1}^{2}-u_{2}^{2}=k^{2}$\,.
The conserved charges for a period become
\begin{equation}
{\cal E} = \f{\pi}{k} \sqrt{M_1^2 + \pare{k^2 - 1} M_2^2} \,, \qquad
{\cal J}_1 = {\cal J}_2 = 0 \,.
\label{charges stat-c}
\end{equation}
Right of Figure \ref{fig:pulsating} shows the time evolution of
the type $(ii)'$ pulsating string. Again, when we set $u_{2}=0$\,,
this string reduces to the simplest pulsating solution studied in
\cite{Minahan:2002rc}.

\subsubsection*{$\bullet$ \bmt{k\to 1} limit\,:~ Single-spike strings}

This limit results in essentially the same solution as the type
$(i)'$ case, that is an array of single-spike strings. The only
difference is that while in the type $(i)'$ case every cusp
appears in the same side about the equator, say the northern
hemisphere, in the type $(ii)'$ case cusps appear in both the
northern and southern hemispheres in turn, each after an infinite
winding.


\subsubsection*{$\bullet$ \bmt{k\to 0} limit\,:~ Rational circular strings}

In the $k\to 0$ limit, the profile becomes
\begin{align}
\hat \eta_{0} = \sqrt{\hat a^2 - \hat b^2} \; \tilde \tau\,,\quad
\hat \xi_{1} = \hat C \; \sin (T - \iomm{1}) \, e^{i u_1 X}\,,\quad
\hat \xi_{2} = \hat C \; \cos (T - \iomm{2}) \, e^{i u_2 X}\,,
\end{align}
with $\hat C = \pare{ \cosh^2 \omega_2 + \sinh^2 \omega_1
}^{-1/2}$ and $u_1^2 = u_2^2 = U + 1$\,. Virasoro constraints imply the following set of relations
between the parameters $\hat a$ and $\hat b$ (with $\hat a \ge
\hat b$):
\begin{align}
\hat a^2 + \hat b^2 &= - U + 2 u_2^2  \,,\\
\hat a \, \hat b &= \hat C^2 \sqrt{U + 1}
\pare{\sinh\omega_1 \cosh \omega_1 \mp \sinh\omega_2 \cosh \omega_2 } \,.
\label{pm}
\end{align}
Here $\mp$ reflects the sign ambiguity in the angular momenta. The
periodicity conditions become
\begin{alignat}{2}
\Delta \sigma &\equiv \frac{2 \pi}{m} = \frac{\pi \sqrt{1 - \hat v^2}}{\hat v \ssp \mu} \,, \\[2mm]
\Delta \varphi_1 &\equiv \frac{2 \pi M_1}{m} = \frac{\pi \ssp u_1}{\hat v} + \pare{2 \ssp m'_1 + 1} \pi \,, \\[2mm]
\Delta \varphi_2 &\equiv \frac{2\pi M_2}{m} = \frac{\pi \ssp u_2}{\hat v} + \pare{2 \ssp m'_2 + 1} \pi \,.
\end{alignat}
The conserved charges for a single period are evaluated as
\begin{equation}
\hat {\cal E} = \frac{\pi \hat a \pare{1 - \hat  v^2}}{2 \ssp \hat v}  \,, \qquad
\hat {\cal J}_1  = \frac{\pi \ssp \hat C^2}{2 \hat v} \, \sinh \omega_1 \cosh \omega_1 \,, \qquad
\hat {\cal J}_2  = - \frac{\pi \ssp \hat C^2}{2 \hat v} \, \sinh \omega_2 \cosh \omega_2 \,.
\end{equation}

\subsubsection*{$\bullet$ \bmt{u_2\,, \omega_2 \to 0}\,: Single-spin limit}

As in the type $(i)'$ case, we obtain the type $(ii)'$ helical
strings with $J_2 = M_2 = 0$ by setting $u_2 = \omega_2 =
0$.\footnote{\,For the type $(ii)'$ case, the other single-spin
limit $u_1 = \omega_1 = 0$ results in $U = - 1$\,, $u_2^2 = - 1 +
(1-k^2)/\dn^2 (\iomm{2})$ and $\hat C = \dn (\iomm{2})/\cn
(\iomm{2})$\,. It turns out equivalent to the $\om_{1,2}\to 0$
limit, because $u_2$ must be real, and thus
the second condition implies $\omega_2 = 0$\,.} Then we find $U =
- 1 + k^2$\,, $u_1 = k \cn (\iom)$ and $\hat C = 1/\cn (\iom)$\,.
The Virasoro conditions require $\hat a = \dn (\iom)$\,, $\hat b =
- i k \sn (\iom)$ and $\hat v = - i k \sn (\iom)/\dn (\iom)$\,.
The periodicity conditions become
\begin{alignat}{3}
\Delta \sigma &= \ \frac{2 \pi}{m} & &= \frac{2 i \eK}{\mu k \sn(\iom)} \,,\qquad  \frac{2\pi M_2}{m} = 0\,,  \\[2mm]
\Delta \varphi_1 &= \frac{2 \pi M_1}{m} & &= 2 i \eK \pare{ \frac{\cn (\iom) \dn (\iom)}{\sn (\iom)} + Z_0 (\iom) } + \pare{2 \ssp m'_1 + 1} \pi  \,,
\label{closed single phi-ii}
\end{alignat}
and the conserved charges for a single period are given by
\begin{equation}
\hat {\cal E} = \frac{i}{k \sn(\iom)} \, \eK \,,\qquad
\hat {\cal J}_1 = \frac{1}{k \cn (\iom)} \, \eE \,,\qquad
\hat {\cal J}_2 = 0\,.
\end{equation}

\section{Finite-gap Interpretation\label{sec:FG}}

\begin{figure}[htbp]
\begin{center}
\vspace{0.5cm}
\includegraphics{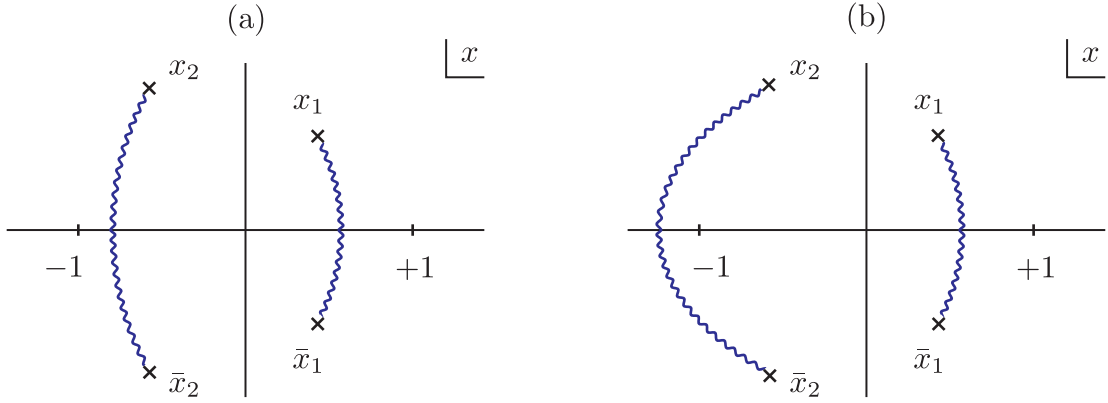}
\vspace{0.5cm}
\caption{\small Different possible arrangements of cuts relative to $x
= \pm 1$\,: (a) corresponds to the helical string, (b) corresponds to
the $\ts$ transformed helical string.}
\label{fig:2cuts}
\end{center}
\end{figure}

The helical strings \eqref{zf1}, \eqref{zf2} of
\cite{Okamura:2006zv} were shown in \cite{Vicedo:2007rp} to be
equivalent to the most general elliptic (``two-cut'') finite-gap
solution on $\mathbb R\times S^{3}\subset AdS_{5}\times S^{5}$\,,
with both cuts intersecting the real axis within the interval
$(-1,1)$ (see Figure \ref{fig:2cuts} (a)). The aim of this
section is to present the corresponding finite-gap description of
the $\ts$ transformed helical string \eqref{zf1-T}, \eqref{zf2-T}
obtained in the previous section.

Recall first from \cite{Vicedo:2007rp} that the $(\sigma,
\tau)$-dependence of the general finite-gap solution enters solely
through the differential form
\begin{equation} \label{dQ}
d \mathcal{Q}(\sigma, \tau) = \frac{1}{2 \pi} \left( \sigma dp +
\tau dq \right)\,,
\end{equation}
where $dp$ and $dq$ are the differentials of the quasi-momentum and
quasi-energy defined below by their respective asymptotics near the
points $x = \pm 1$. The differential multiplying $\sigma$ in
$d\mathcal{Q}(\sigma,\tau)$ (namely $dp$) is related to the eigenvalues of the monodromy
matrix, which by definition is the parallel transporter along
a closed loop $\sigma \in [ 0, 2 \pi]$ on the worldsheet. This is because
the Baker-Akhiezer vector $\bmt{\psi}(P,\sigma,\tau)$, whose $(\sigma,
\tau)$-dependence also enters solely through the differential form $d
\mathcal{Q}(\sigma, \tau)$ in \eqref{dQ}, satisfies \cite{Dorey:2006zj}
\begin{equation*}
\bmt{\psi}(P,\sigma + 2 \pi,\tau) = \exp \left\{ i \int_{\infty^+}^P dp
\right\} \bmt{\psi}(P,\sigma,\tau)\,.
\end{equation*}
Now it is clear from \eqref{dQ} that the $\sigma \leftrightarrow \tau$ operation can be realised on the general
finite-gap solution by simply interchanging the quasi-momentum with
the quasi-energy,
\begin{equation} \label{dp-dq}
dp ~\leftrightarrow~ dq\,.
\end{equation}
However, since we wish $dp$ to always denote the differential related
to the eigenvalues of the monodromy matrix, by the above argument it
must always appear as the coefficient of $\sigma$ in $d
\mathcal{Q}(\sigma, \tau)$. Therefore equation \eqref{dp-dq} should be
interpreted as saying that the respective definitions of the
differentials $dp$ and $dq$ are interchanged, but $d
\mathcal{Q}(\sigma, \tau)$ always takes the same form as in
\eqref{dQ}.

Before proceeding let us recall the precise definitions of these
differentials $dp$ and $dq$\,. Consider an algebraic curve
$\Sigma$\,, which admits a hyperelliptic representation with cuts.
For what follows it will be important to specify the position of
the different cuts relative to the points $x = \pm 1$\,, {\em
i.e.}, Figures \ref{fig:2cuts} (a) and \ref{fig:2cuts} (b) are to
be distinguished for the purpose of defining $dp$ and $dq$\,. We
could make this distinction by specifying an equivalence relation
on representations of $\Sigma$ in terms of cuts, where two
representations are equivalent if the cuts of one can be deformed
into the cuts of the other within $\mathbb{C} \setminus \{ \pm 1
\}$\,. It is straightforward to see that there are only two such
equivalence classes for a general algebraic curve $\Sigma$\,. For
example, in the case of an elliptic curve $\Sigma$ the
representatives of these two equivalence classes are given in
Figures \ref{fig:2cuts} (a) and \ref{fig:2cuts} (b). Now with
respect to a given equivalence class of cuts, the differentials
$dp$ and $dq$ can be uniquely defined on $\Sigma$ as in
\cite{Dorey:2006zj} by the following conditions:
\begin{itemize}
\item[$(1)$] their $\mathcal A$-period vanishes.
\item[$(2)$] their respective poles at $x = \pm
1$ are of the following form, up to a trivial overall change of sign
(see \cite{Vicedo:2007rp}),
\begin{equation} \label{dp asymptotics at pm 1}
dp(x^{\pm}) \underset{x \rightarrow +1}\sim \mp \frac{\pi \kappa dx}{(x -
1)^2}\,, \qquad dp(x^{\pm}) \underset{x \rightarrow -1}\sim \mp
\frac{\pi \kappa dx}{(x + 1)^2}\,,
\end{equation}
\begin{equation} \label{dq asymptotics at pm 1}
dq(x^{\pm}) \underset{x \rightarrow +1}\sim \mp \frac{\pi \kappa dx}{(x -
1)^2}\,, \qquad dq(x^{\pm}) \underset{x \rightarrow -1}\sim \pm
\frac{\pi \kappa dx}{(x + 1)^2}\,,
\end{equation}
where $x^{\pm} \in \Sigma$ denotes the pair of points above $x$\,, with
$x^{+}$ being on the physical sheet, and $x^{-}$ on the other sheet.\footnote{\,
They should not be confused with AdS/CFT spectral parameters (\ref{xpm}).
}
\end{itemize}
Once the differentials $dp$ and $dq$ have been defined by \eqref{dp
asymptotics at pm 1} and \eqref{dq asymptotics at pm 1} with respect
to a given equivalence class of cuts, one can move the cuts around
into the other equivalence class (by crossing say $x = - 1$ with a
single cut) to obtain a representation of $dp$ and $dq$ with respect
to the other equivalence class of cuts. So for instance, if we
define $dp$ and $dq$ by \eqref{dp asymptotics at pm 1} and \eqref{dq
asymptotics at pm 1} with respect to the equivalence class of cuts
in Figure \ref{fig:2cuts} (a), then with respect to the equivalence
class of cuts in Figure \ref{fig:2cuts} (b) the definition of $dp$
will now be \eqref{dq asymptotics at pm 1} and that of $dq$ will now
be \eqref{dp asymptotics at pm 1}.

In summary, both equivalence classes of cuts represents the very
same algebraic curve $\Sigma$\,, but each equivalence class gives
rise to a different definition of $dp$ and $dq$\,. So the two
equivalence classes of cuts give rise to two separate finite-gap
solutions but which can be related by a $\ts$ transformation
\eqref{dp-dq}. Indeed, if in the construction of
\cite{Vicedo:2007rp} we assume the generic configuration of cuts
given in Figure \ref{fig:2cuts} (b), instead of Figure
\ref{fig:2cuts} (a) as was assumed in \cite{Vicedo:2007rp}, then
the resulting solution is the generic helical string but with
\begin{equation*}
X ~\leftrightarrow~ T
\end{equation*}
namely the 2D transformed helical string \eqref{zf1-T},
\eqref{zf2-T}. Therefore, with $dp$ and $dq$ defined as above by their
respective asymptotics \eqref{dp asymptotics at pm 1} and \eqref{dq
asymptotics at pm 1} at $x = \pm 1$, the helical string of \cite{Okamura:2006zv,
Vicedo:2007rp} is the general finite-gap solution corresponding to the
class represented by Figure \ref{fig:2cuts} (a), whereas the
2D transformed helical string corresponds to the most general elliptic
finite-gap solution on $\mathbb R\times S^{3}$ with cuts in the other
class represented in Figure \ref{fig:2cuts} (b).

As is clear from the above, a given finite-gap solution is not
associated with a particular equivalence class of cuts; since $dp$ and $dq$ are
defined relative to an equivalence class of cuts, one can freely change
equivalence class provided one also changes the definitions of $dp$
and $dq$ with respect to this new equivalence class according to
\eqref{dp-dq}, so that in the end $dp$ and $dq$ define the same
differentials on $\Sigma$ in either representation. For example, we
can describe the 2D transformed helical string in two different ways:
either we take the  configuration of cuts in Figure \ref{fig:2cuts}
(b) with $dp$ and $dq$ defined as usual by their asymptotics \eqref{dp
asymptotics at pm 1} and \eqref{dq asymptotics at pm 1} at $x = \pm
1$\,, or we take the configuration of cuts in Figure \ref{fig:2cuts}
(a) but need to swap the definitions of $dp$ and $dq$ in \eqref{dp
asymptotics at pm 1} and \eqref{dq asymptotics at pm 1}. In the
following we will use the latter description of Figure \ref{fig:2cuts}
(a) in order to take the singular limit $k \to 1$ where the cuts merge
into a pair of singular points.

We can obtain expressions for the global charges $J_1
=(J_{L}+J_{R})/2$\,, $J_2 = (J_{L}-J_{R})/2$ along the same lines as
in \cite{Vicedo:2007rp} for the helical string. In terms of the
differential form
\begin{equation}
\alpha \equiv \frac{\sqrt{\lambda}}{4 \pi} \left( x + \frac{1}{x}
\right) dp\,, \qquad \tilde{\alpha} \equiv \frac{\sqrt{\lambda}}{4
\pi}\left( x - \frac{1}{x} \right) dp\,,
\end{equation}
we can write
\begin{align}
J_1 &= -\text{Res}_{0^+} \alpha + \text{Res}_{\infty^+} \alpha =
\text{Res}_{0^+} \tilde{\alpha} + \text{Res}_{\infty^+}
\tilde{\alpha}\,, \label{J_1 res}\\
J_2 &= -\text{Res}_{0^+} \alpha - \text{Res}_{\infty^+} \alpha\,. \label{J_2 res}
\end{align}
Note that $\alpha$ and $\tilde{\alpha}$ both have simple poles at $x =
0$\,, $\infty$ but $\tilde{\alpha}$ also has simple poles at $x = \pm 1$
coming from the double poles in $dp$ at $x = \pm 1$\,. It follows that
we can rewrite \eqref{J_1 res}, \eqref{J_2 res} as
\begin{align}
J_1 &= - \sum_{I=1}^2 \frac{1}{2 \pi i} \int_{{\mathcal A}_I} \tilde{\alpha} -
\text{Res}_{(+1)^+} \tilde{\alpha} - \text{Res}_{(-1)^+} \tilde{\alpha}\,,\\
J_2 &= \sum_{I=1}^2 \frac{1}{2 \pi i} \int_{{\mathcal A}_I} \alpha\,,
\end{align}
where ${\mathcal A}_I$ is the $\mathcal A$-cycle around the $I$-th cut.
Whereas in \cite{Vicedo:2007rp} the residues of $\tilde{\alpha}$ at $x
= \pm 1$ were of the same sign (as a consequence of $p(x)$ having
equal residues at $x = \pm 1$) so that their sum gave the energy $E$
of the string, in the present 2D-transformed helical case the residues of
$\tilde{\alpha}$ at $x = \pm 1$ are now opposite (since $p(x)$ now has
opposite residues at $x = \pm 1$) and therefore cancel in the above
expression for $J_1$\,, resulting in the following expressions
\begin{equation} \label{J_1,J_2 periods}
- J_1 = \sum_{I=1}^2 \frac{1}{2 \pi i} \int_{{\mathcal A}_I} \tilde{\alpha}\,,
\qquad J_2 = \sum_{I=1}^2 \frac{1}{2 \pi i} \int_{{\mathcal A}_I} \alpha\,.
\end{equation}

\begin{figure}[htbp]
\begin{minipage}{0.5\hsize}
\begin{center}
\vspace{0.5cm}
\includegraphics{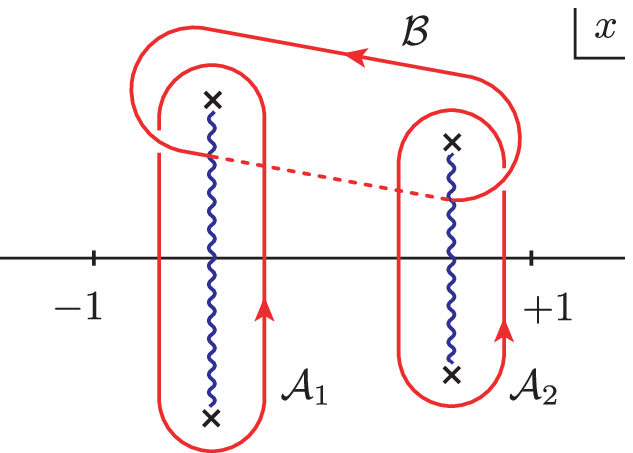}
\vspace{0.5cm}
\caption{\small Definitions of cycles.}
\label{fig:cycles}
\end{center}
\end{minipage}
\begin{minipage}{0.5\hsize}
\begin{center}
\vspace{0.5cm}
\includegraphics{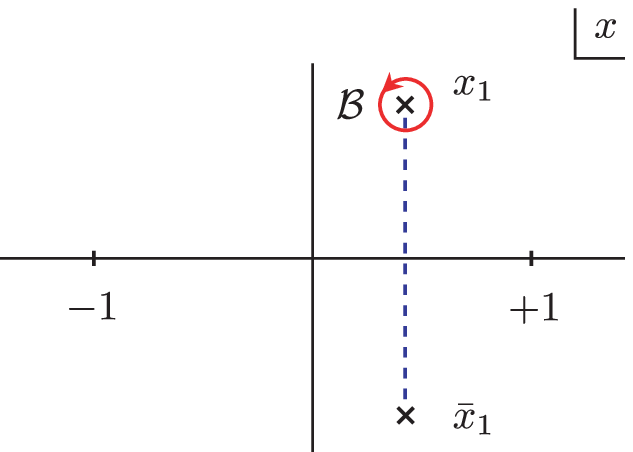}
\vspace{0.5cm}
\caption{\small $k\to 1$ limit of cuts.}
\label{fig:ss-FG}
\end{center}
\end{minipage}
\vspace{0.5cm}
\end{figure}

\paragraph{}
In parallel to the discussion of the helical string case in \cite{Vicedo:2007rp}, there are two types of limits one can consider: the symmetric cut limit (where the curve acquires the extra symmetry $x \leftrightarrow -x$) which corresponds to taking $\omega_{1,2} \rightarrow 0$ in the finite-gap solution, or the singular curve limit which corresponds to taking the moduli of the curve to one, $k \rightarrow 1$\,. In the symmetric cut limit the discussion is identical to that in \cite{Vicedo:2007rp} (when working with the configuration of cuts in Figure \ref{fig:2cuts} (a)), in particular there are two possibilities corresponding to the type $(i)'$ and type $(ii)'$ cases, for which the cuts are symmetric with $x_1 = -\bar{x}_2$ and imaginary with $x_1 = -\bar{x}_1$\,, $x_2 = -\bar{x}_2$ respectively (see Figure 2 of \cite{Vicedo:2007rp}).

\paragraph{}
In the singular limit $k \to 1$ where both cuts merge into a
pair of singular points at $x = x_1$\,, $\bar{x}_1$ \cite{Vicedo:2007rp}, the
sum of $\mathcal A$-cycles turns into a sum of cycles around the points $x_1$\,, $\bar{x}_1$\,, so that \eqref{J_1,J_2 periods} yields in this limit
\begin{equation} \label{J_1,J_2 res}
- J_1 = \text{Res}_{x_1} \tilde{\alpha} + \overline{\text{Res}_{x_1}
\tilde{\alpha}}\,, \qquad
J_2 = \text{Res}_{x_1} \alpha + \overline{\text{Res}_{x_1} \alpha}\,.
\end{equation}
Moreover, in the singular limit $dp$ acquires simple poles at $x =
x_1$\,, $\bar{x}_1$ so that the periodicity condition about the $\mathcal B$-cycle, $\int_{\mathcal B} dp = 2 \pi n$\,, implies
\begin{equation*}
\text{Res}_{x_1} dp = \frac{n}{i}\,.
\end{equation*}
Let us set $n=1$ ($n$ can be easily recovered at any moment).
Then \eqref{J_1,J_2 res} simplifies to
\begin{align}
- J_1 &= \frac{\sqrt{\lambda}}{4 \pi} \left| \left( x_1 -
\frac{1}{x_1} \right) -\left( \bar{x}_1 - \frac{1}{\bar{x}_1}
\right) \right|\,, \label{J_1 res-2}\\ J_2 &= \frac{\sqrt{\lambda}}{4 \pi}
\left| \left( x_1 + \frac{1}{x_1} \right) - \left( \bar{x}_1 +
\frac{1}{\bar{x}_1} \right) \right|\,.\label{J_2 res-2}
\end{align}
The energy $E = \sqrt{\lambda}\,\kappa = (n \sqrt{\lambda}/\pi)
\,\mathcal{E}$ diverges in the singular limit $k \to 1$\,, but
this divergence can be related to the one in $\Delta \varphi_1$\,.
In the present case the $\sigma$-periodicity condition
$\int_{\mathcal B} dp \in 2 \pi \mathbb{Z}$ can be written as
({\em c.f.}, equation $(2.23)$ in \cite{Vicedo:2007rp})
\begin{equation*}
- \frac{2 \eK \sqrt{1 - v^2}}{v} = \frac{2 \pi}{n} \kappa' \equiv \frac{2 \pi
\kappa |x_1 - \bar{x}_2|}{n \sqrt{y_+ y_-}}\,,
\end{equation*}
where $\eK = \eK(k)$\,, $y_{\pm} = y(x)|_{x = \pm 1} > 0$\,, $y(x)=(x-x_{1})(x-\bar{x}_{1})(x-x_{2})(x-\bar{x}_{2})$ and $v$ can be
expressed in the present setup as $v = \frac{y_+ - y_-}{y_+ + y_-}$
(see \cite{Vicedo:2007rp}). Using this $\sigma$-periodicity condition
the energy can be expressed in the $k \to 1$ limit as
\begin{equation*}
\mathcal{E} = \frac{u_1}{v} (1 - v^2) \,\eK(1)\,.
\end{equation*}
We can relate this divergent expression with the expression
\eqref{Dphi1_cl f} for $\Delta \varphi_1$ which also diverge in the
limit $k \to 1$\,, making use of the relation $u_1 v = \tan \omega_1$ (see
\cite{Vicedo:2007rp} where the notation is $u_1 = v_-$ and $\omega_1 =
\tilde{\rho}_-$), and find
\begin{equation}
\mathcal{E} - \frac{\Delta \varphi_1}{2} = - \left( \omega_1 -
\frac{(2 n'_1 + 1) \pi}{2} \right) \equiv \bar{\theta}\,. \label{E fg}
\end{equation}
Comparing this scenario with the one for helical strings in
\cite{Vicedo:2007rp} we can write an expression for $\bar{\theta}$ in
terms of the spectral data $x_1$ of the singular curve. Identifying
\begin{equation}
\bar{\theta} = -\frac{i}{2} \ln
\left(\frac{x_1}{\bar{x}_1}\right), \label{theta fg}
\end{equation}
the expressions \eqref{J_1 res-2}, \eqref{J_2 res-2} and \eqref{theta fg} together
imply the relation\footnote{\,
The sign difference between (\ref{spin spin}) and here is not essential.}
\begin{equation}
-J_1 = \sqrt{J_2^2 + \frac{\lambda}{\pi^2} \sin^2
 \bar{\theta}}\,.
\end{equation}

\section{Gauge Theory Duals\label{sec:gauge theory}}

In view of the pulsating (oscillating) nature of the $\ts$ transformed helical strings we saw in the previous sections, the gauge theory operators dual to those classical strings should be made up not only of holomorphic but also of non-holomorphic scalars.
In this section we discuss the gauge theory interpretation of 2D transformed strings, which includes a single-spike string and a static circular string.

\paragraph{}
First let us review some relevant aspects of the $SU(2)$ magnon boundstates.
Let $Z$ or $W$ be two of the three complex scalar fields of $\mathcal N=4$ SYM (the third one will be denoted $Y$).
Then operators in the $SU(2)$ sector take the forms $\mathcal O=\tr\ko{\Phi_{i_{1}}\Phi_{i_{2}}\dots}+\dots$ with each $\Phi_{i_{l}}$ $(l=1,\dots,L)$ being either $Z$ or $W$\,.
The BPS operator $\tr\ko{ZZ\dots}$ made up only of $Z$ is the ferromagnetic ground state for the SYM spin-chain.
In \cite{Chen:2006ge}, it is shown that dyonic giant magnons are dual to magnon boundstates $\mathcal O_{\rm DGM}\sim \tr\ko{Z^{K}W^{M}}+\dots$ in the SYM spin-chain ($K\to \infty$\,, $M$: finite), whose dispersion relation is given by 
\begin{equation}
\Delta_{\mathcal O_{\rm DGM}}-K=\sqrt{M^{2}+16g^{2}\sin^{2}\ko{\f{P}{2}}}\,,
\qquad
g\eq \f{\sqrt{\lam}}{4\pi}\,.
\end{equation}
This agrees with the energy-spin relation for a dyonic giant magnon under the identifications $J_{1}=K~(\to \infty)$ and $J_{2}=M$\,.
Here $P=\sum_{j=1}^{M}p_{j}$ is the sum of the momenta $p_{j}$ $(j=1,\dots, M)$ of the constituent magnons.
They satisfy the following boundstate condition,
\begin{equation}
x^{-}(p_{j}) = x^{+}(p_{j+1})
\qquad \mbox{for}\quad 
j=\komoji{1,\dots, M}\,,
\end{equation}
where $x^{\pm}(p)$ are the standard AdS/CFT spectral parameters, defined by
\begin{equation}
x^{\pm}(u)=x\ko{u\pm\f{i}{2g}}
\quad \mbox{where}\quad 
x(u)=\f{1}{2}\ko{u+\sqrt{u^{2}-4}}\,,
\label{xpm}
\end{equation}
and $u(p)$ is the rapidity variable,
\begin{equation}
u(p)=\f{1}{2}\cot\ko{\f{p}{2}}\sqrt{1+16g^{2}\sin^{2}\ko{\f{p}{2}}}\,.
\end{equation}

\paragraph{}
Now let us turn to the present oscillating case.
First we discuss the two-spin single-spike string case. As we
have seen, in contrast to the dyonic giant magnon, it has finite spins $J_{i}$ $(i=1,2)$ and infinite
energy. This fact allows us to claim that the relevant dual SYM
operators should look like
\begin{equation}
\mathcal O_{\rm SS}= \tr\ko{Z^{K}\,\overline{Z}{}^{K'}\,W^{M}\,{\mathcal S}^{(L-K-K'-M)/2}}+\dots\,,\quad
L\,, K\,, K'\to \infty\,,\quad K-K'\,, M : \mbox{finite}\,.
\label{op}
\end{equation}
In (\ref{op}), the factor
${\mathcal S}$ appearing in (\ref{op}) is the $SO(6)$\,-singlet
composite\footnote{\, The $SO(6)$ sector is not closed beyond
one-loop level in $\lam$\,, and operator mixing occurs in the full
$PSU(2,2|4)$ sector due to the higher-loop effects. So one might
think $\mathcal S$ should be a $PSU(2,2|4)$ singlet rather than an
$SO(6)$ singlet. However, we can still expect that such mixing
into $PSU(2,2|4)$ is suppressed in our classical ($L\to \infty$)
setup as in \cite{Minahan:2004ds}. We would like to thank J.~Minahan for discussing this
point. }
\begin{equation}
\mathcal S\sim Z\overline{Z}+W\overline{W}+Y\overline{Y}\,.
\label{SO(6) singlet}
\end{equation}
One can easily understand that the pairs like $Z\overline{Z}$ give rise to oscillating motion in the sting side, since if we associate $Z$ to a particle rotating along a great circle of $S^{5}$ clockwise, the other particle associated with $\overline{Z}$ rotates counterclockwise, thus making the string connecting these two points non-rigid and oscillating.
The dots in (\ref{op}) denotes terms that mix under
renormalization. An important assumption is that $M$ $W$s form a
boundstate. Indeed loop-effects mix $Z\overline{Z}$ with other
neutral combinations $W\overline{W}$ and $Y\overline{Y}$\,, but it is assumed the boundstate condition still holds. 
Let $X^{\pm}$ be the spectral parameters assigned to the
boundstate. We write them as
\begin{alignat}{3}
X^{\pm}&=R\,e^{\pm iP/2}
&\quad &\mbox{with}\quad R=\f{M+\sqrt{M^{2}+16g^{2}\sin^{2}\ko{P/2}}}{4g\sin\ko{P/2}} ~ (>1)\,,
\end{alignat}
where $P$ is the momentum carried by the boundstate. Recall that
we took $\tr\ko{ZZ\dots}$ as the vacuum state, therefore $W$ is an
excitation above the vacuum with
$\Delta_{0}-J_{1}=1$\,,\footnote{\, We follow a convention such
that a $Z$ field has $\Delta_{0}-J_{1}=0$\,, where $\Delta_{0}$
denotes the bare dimension. } whereas $\overline{Z}$ is an
excitation with $\Delta_{0}-J_{1}=2$\,.\footnote{\,
In fact, $\overline{Z}$ is not a fundamental excitation. We should regard it as an excitation corresponding to a two-magnon state.} The composite ${\mathcal
S}$ also contributes to the spin-chain energy in some way, and we
must take all the contributions into account when evaluating the
total energy $\Delta_{\mathcal O_{\rm SS}}-J_{1}$ of (\ref{op}).
We assume that the contribution of $M$ $W$s results in two parts;
one is the boundstate energy that contributes in the same way as
in the case of an $SU(2)$ boundstate $\mathcal O_{\rm DGM}\sim
\tr\ko{Z^{K}W^{M}}+\dots$ $(K\to \infty)$\,, and the other is its
interactions with other fields. One can then write down the total
energy as
\begin{align}
\Delta_{\mathcal O_{\rm SS}}-(K-K')=
\f{g}{i}\kko{\ko{X^{+}-\f{1}{X^{+}}}-\ko{X^{-}-\f{1}{X^{-}}}}
+\chi\,.
\label{dr1}
\end{align}
The first term in RHS comes from the boundstate $W^{M}$\,, while the last $\chi$ accounts for contributions concerning ${\mathcal S}$\,, $\overline{Z}$ and all their interactions with other fields, including $W$s\,. Currently we have no
knowledge of how the actual form of $\chi$ looks like, and so we leave it as some function of the coupling and boundstate momentum here (however, we will
later discuss its form in the strong coupling, infinite-winding
limit). One can also express the $J_{2}$\,-charge carried by the
boundstate in terms of the spectral parameters as
\begin{align}
M=\f{g}{i}\kko{\ko{X^{+}+\f{1}{X^{+}}}-\ko{X^{-}+\f{1}{X^{-}}}}\,.
\label{M1}
\end{align}

Now perform a change of basis for the spin-chain, and take $\tr\ko{\overline{Z}\,\overline{Z}\dots}$ as the vacuum state, instead of $\tr\ko{ZZ\dots}$\,.
This particular transformation of susy multiplet, namely the charge conjugation, maps the original $W^{M}$ to $\overline{W}{}^{M}$ with new spectral parameters
\begin{equation}
{\widetilde X}^{\pm}=1/X^{\pm}\,.
\end{equation}
This is actually a crossing transformation that maps a usual
particle to its conjugate particle (antiparticle)
\cite{Beisert:2005tm}. In the new basis, $\overline{W}$s, $Z$s and
$\overline {\mathcal S}={\mathcal S}$ play the role of excitations
above the new vacuum. The contribution of $\mathcal S$ to the new vacuum should be the same as in the old case since it is an $SO(6)$ singlet, and we assume the total contributions from all excitations to be the same as in the old case.
Then one obtains a relation similar to
(\ref{dr1}),
\begin{align}
\Delta_{\mathcal O_{\rm SS}}-(K'-K)=\f{g}{i}\kko{\ko{{\widetilde X}^{+}-\f{1}{{\widetilde X}^{+}}}-\ko{{\widetilde X}^{-}-\f{1}{{\widetilde X}^{-}}}}+\chi\,,
\label{dr2}
\end{align}
and similarly for the second charge. From (\ref{dr1})-(\ref{dr2}),
it follows that
\begin{align}
\Delta_{\mathcal O_{\rm SS}}=\chi
\qquad\mbox{and}\qquad
K'-K=\sqrt{M^{2}+16 g^{2}\sin^{2}\ko{\f{P}{2}}}\,.\label{Delta and K-K'}
\end{align}
Then if we identify naturally
\begin{equation}
K-K'\eq J_{1}\,,\quad
M\eq J_{2}\quad
\mbox{and}\quad
P\eq 2\pi m \pm 2\bar \theta\quad (m\in \mathbb Z\,; ~ 0\leq \bar{\theta}\leq \pi/2)\,,
\label{identification1}
\end{equation}
the second relation in (\ref{Delta and K-K'}) precisely reproduces
the dispersion relation for single-spike strings, after
substituting $g^{2}=\lam/16\pi^{2}$\,. Here we included an integer
degree of freedom $m$ that plays the role of the winding number in
the string theory side. One can also deduce that
\begin{equation}
\f{J_{2}}{J_{1}}=\f{R^{2}-1}{R^{2}+1}\,,
\end{equation}
which corresponds to $\sin\gamma$ in the notation used in
\cite{Ishizeki:2007we}. In (\ref{identification1}), one may choose
either the plus/minus signs in $P$\,; they correspond to the
momenta of a particle/antiparticle.

Notice also the above argument, resulting in
\begin{align}
-J_{1}&=\f{g}{i}\kko{\ko{X^{+}-\f{1}{X^{+}}}-\ko{X^{-}-\f{1}{X^{-}}}}\,,\\
J_{2}&=\f{g}{i}\kko{\ko{X^{+}+\f{1}{X^{+}}}-\ko{X^{-}+\f{1}{X^{-}}}}\,,
\end{align}
is consistent with what we found in the previous section, (\ref{J_1 res-2}) and (\ref{J_2 res-2}), if we, as usual, identify the string theory spectral parameters $x_{1}$ and $\bar{x}_{1}$ (in finite-gap language) with the ones for gauge theory $X^{+}$ and $X^{-}$ (for the boundstate).

\paragraph{}
To proceed in the reasoning, suppose the asymptotic behavior of
$\chi$ in the strong coupling and infinite-``winding'' limit
becomes
\begin{equation}
\chi\sim 2gP=m\sqrt{\lam}\pm \f{\bar \theta}{\pi}\,,\qquad (m\to \infty)\,.
\label{c}
\end{equation}
We kept here $\pm\bar\theta/\pi$ term to ensure that $\chi$ is not
just given by $\mbox{(integer)}\times \sqrt{\lam}$ but contains
some continuous shift away from that. We will give more
explanations concerning this conjecture soon. The relation
(\ref{c}) then implies that
\begin{equation}
\Delta_{\mathcal O_{\rm SS}} - \f{\sqrt{\lam}}{2\pi}\cdot 2\pi m
=\pm\f{\sqrt{\lam}}{\pi}\,\bar\theta\,,
\label{ene-wind-2}
\end{equation}
where we used the identifications we made before. This can be
compared to the string theory result for the single-spike,
(\ref{ene-wind}). The integer $m$ here corresponds to the winding
number $N_{1}$ there (recall that for single spike case, we had
$\Delta \varphi_{1}=2\pi N_{1}$ due to the periodicity condition).
When there are $n$ boundstates
in the spin-chain all with the same momentum $P$\,, RHS of
(\ref{ene-wind-2}) is just multiplied by $n$ and modified to
$n(\sqrt{\lam}/\pi)\,\bar \theta$\,, which corresponds to an array
of $n$ single-spikes.

\paragraph{}
Let us explain the conjecture (\ref{c}) in greater detail. Of
course one of the motivations is that it reproduces the relation
(\ref{ene-wind-2}) of the string side, as we have just seen. Further
evidence can be found by considering particular sets of operators
contained in (\ref{op}) and checking for consistency.
For example, let us consider the limit $K-K'\to 0$ and $M\to 0$\,.
This takes the operator (\ref{op}) to the form $\tr\ko{(Z\overline{Z})^{K}{\mathcal
S}^{L/2-K}}+\dots$\,, which must sum up to the singlet operator $\tr{\mathcal
S}^{L/2}$ for it to be a solution of the Bethe ansatz equation.
In this limit, the ``angle'' $\bar \theta$ should vanish in view of the second
equation in (\ref{Delta and K-K'}) and (\ref{identification1}).
Therefore the relation (\ref{c}) together with the first equation
in (\ref{Delta and K-K'}) imply that the canonical dimension of
the singlet operator is just given by
\begin{equation}
\Delta_{\tr{\mathcal S^{L/2}}}\big|_{L\to \infty}=m\sqrt{\lam}\,,\qquad (m\to \infty)\,,
\label{k->1 pulsating}
\end{equation}
which agrees with the energy expression (\ref{winding energy}) of the $\ts$ transformed point-like BPS string (in the limit $\mu\sqrt{U}\to \infty$), under the identification $N_{2}=m$\,.

\paragraph{}
As we have seen, in contrast to the dyonic giant magnon vs.\
magnon bound state $\mathcal O_{\rm DGM}\sim
\tr\ko{Z^{\infty}W^{M}}+\dots$ case, the correspondence between two-spin
single-spike vs.\ $\mathcal O_{\rm SS}$ given in (\ref{op}) is
slightly more involved. In the former correspondence in the
infinite spin sector, the magnon boundstate is an excitation above
the BPS vacuum $\mathcal O_{\rm F}\sim \tr\ko{Z^{\infty}}$\,, and
one can think of the boundstate $W^{M}$ as the counterpart of the
corresponding dyonic giant magnon. For the latter case in the
infinite winding sector, however, it is not the boundstate $W^{M}$
alone but the ``$Z^{K}\,\overline{Z}{}^{K'}\,W^{M}+\dots$'' part of
$\mathcal O_{\rm SS}$ that encodes the single-spike. It can be
viewed as an excitation above the $SO(6)$ singlet operator
$\mathcal O_{\rm AF}\sim \tr \mathcal S^{L/2}$\,. Actually this is the
``antiferromagnetic'' state of the $SO(6)$ spin-chain, which is
``the farthest from BPS'' (Notice that a solution of the Bethe ansatz equation with
$J_{1}=J_{2}=J_{3}=0$ is nothing but the $SO(6)$ singlet state).
It is dual to the rational circular static string (\ref{rational static
prof}) obtained by performing a $\tau\leftrightarrow\sig$
transformation on the point-like BPS string.

\section{Summary and Discussions\label{sec:summary}}

In the previous works \cite{Okamura:2006zv,Vicedo:2007rp}, three
of the current authors constructed the most general elliptic
(``two-cut'') classical string solutions on $\mathbb R\times
S^{3}\subset AdS_{5}\times S^{5}$\,, called helical strings. They
were shown to include various strings studied in the {\em
large-spin} sector. Schematically, the family tree reads
\begin{alignat}{5}
&\, {\rm I}&{}:\quad &{}
\begin{array}{l}
\mbox{Type $(i)$ helical string} \\
\mbox{with generic $k$ and $\om_{1,2}$}
\end{array}
~ &\longrightarrow~ &
\left\{
\begin{array}{ll}
\mbox{- Point-like (BPS), rotating string} & (k\to 0) \\
\mbox{- Array of dyonic giant magnons} & (k\to 1) \\
\mbox{- Elliptic, spinning folded string} & (\om_{1,2}\to 0)
\end{array}
\right.\,,\no\\[2mm]
&{\rm II}&{}:\quad &{}
\begin{array}{l}
\mbox{Type $(ii)$ helical string} \\
\mbox{with generic $k$ and $\om_{1,2}$}
\end{array}
~ &\longrightarrow~ &
\left\{
\begin{array}{ll}
\mbox{- Rational, spinning circular string} & (k\to 0) \\
\mbox{- Array of dyonic giant magnons} & (k\to 1) \\
\mbox{- Elliptic, spinning circular string} & (\om_{1,2}\to 0)
\end{array}
\right.\,.\no
\end{alignat}
Moreover, the single-spin limit of the type $(i)$ helical strings
agrees with so-called ``spiky strings'' studied in
\cite{Ryang:2005yd, Arutyunov:2006gs-Astolfi:2007uz}.\footnote{\,
The two-spin helical strings are different from the spiky strings
in that they have no singular points in spacetime. When embedded
in $\mathbb R\times S^{3}$\,, the singular ``cusps'' of the spiky
string that apparently existed on $\mathbb R\times S^{2}$ are all
smoothed out to result in non-spiky profiles. }

For Cases ${\rm I}$ and ${\rm II}$\,, the gauge theory duals are
also well-known. They are all of the form
\begin{equation}
\mathcal O\sim \tr\ko{Z^{L-M}W^{M}}+\dots\,,
\end{equation}
 with $L$ very large.
For example, for the type $(i)$ case, a BPS string ($k\to 0$) of
course corresponds to $M=0$\,, and a BMN string corresponds to $M$
very small. A dyonic giant magnon corresponds to an $M$-magnon
boundstate in the asymptotic SYM spin-chain ($L\to \infty$), which
is described by a straight Bethe string in rapidity plane
\cite{Dorey:2006dq, Chen:2006gq}. In the Bethe string, all $M$
roots are equally spaced in the imaginary direction, reflecting
the pole condition of the asymptotic S-matrix. As to the elliptic
folded/circular strings, they correspond to, respectively, the
so-called double-contour/imaginary-root distributions of Bethe
roots \cite{Beisert:2003xu}.

\paragraph{}
In contrast, in the current paper, we explored non-holomorphic
sector of classical strings on $\mathbb R \times S^{3}$\,, and
found a new interpolation. This includes a {\em large-winding}
sector where $m\sqrt{\lam}$ becomes of the same order as the
energy which diverges ($m$ being the winding number). We saw that
when classical strings on $\mathbb R \times S^3 \subset AdS_5
\times S^5$ are considered in conformal gauge, an operation of
interchanging $\tau$ and $\sigma$\,, as well as keeping temporal
gauge $t \propto \tau$\,, maps the original helical strings to
another type of helical strings. Roughly speaking,
rotating/spinning solutions with large spins became oscillating
solution with large windings. Again, schematically, we found\,:
\begin{alignat}{5}
&\, {\rm I}'&{}:\quad &{}
\begin{array}{l}
\mbox{Type $(i)'$ helical string} \\
\mbox{with generic $k$ and $\om_{1,2}$}
\end{array}
~ &\longrightarrow~ &
\left\{
\begin{array}{ll}
\mbox{- Rational, static circular string} & (k\to 0) \\
\mbox{- Array of single-spike strings} & (k\to 1) \\
\mbox{- Elliptic, type $(i)'$ pulsating string} & (\om_{1,2}\to 0)
\end{array}
\right.\,,\no\\[2mm]
&{\rm II}'&{}:\quad &{}
\begin{array}{l}
\mbox{Type $(ii)'$ helical string} \\
\mbox{with generic $k$ and $\om_{1,2}$}
\end{array}
~ &\longrightarrow~ &
\left\{
\begin{array}{ll}
\mbox{- Rational circular string} & (k\to 0) \\
\mbox{- Array of single-spike strings} & (k\to 1) \\
\mbox{- Elliptic, type $(ii)'$ pulsating string} & (\om_{1,2}\to 0)
\end{array}
\right.\,.\no
\end{alignat}

\paragraph{}
In Section \ref{sec:FG}, we investigated 2D-transformed helical
strings from the finite-gap perspective. We were able to
understand the effect of the $\tau \leftrightarrow \sigma$
operation as an interchange of quasi-momentum and quasi-energy.
The transformed helical strings were described as general two-cut
finite-gap solutions as in the original case \cite{Vicedo:2007rp},
the only difference being the asymptotic behaviors of
differentials at $x\to \pm 1$ (or equivalently, different
configurations of cuts with respect to interval $(-1, 1)$). By
expressing the charges in terms of spectral parameters
(branch-points of the cuts), the charge relations for single
spikes were also reproduced.

\paragraph{}

In Section \ref{sec:gauge theory}, the gauge theory duals of the
$\ts$ transformed strings (derivatives of type $(i)'$ and $(ii)'$
helical strings) were identified with operators of the form
\begin{equation}
\mathcal O\sim \tr\ko{Z^{K}\,\overline{Z}{}^{K'}\,W^{M}\,{\mathcal S}^{(L-K-K'-M)/2}}+\dots
\label{op:gen}
\end{equation}
with $\mathcal S$ the $SO(6)$ singlet composite (\ref{SO(6)
singlet}). The single-spike limit $k\to 1$ was identified with the
$K\,, K'\to \infty$ limit while keeping $K-K'$ and $M$ finite (see
(\ref{op})). In this limit, the ``$Z^{K}\,\overline{Z}{}^{K'}\,W^{M}+\dots$''
part in the operator, of which $W^{M}$ is assumed to form a boundstate, was claimed to be responsible for the transverse
excitation (spikes) of the string state winding infinitely many
times around a great circle of $S^{5}$\,. In other words, the
spikes are dual to excitations above the ``antiferromagnetic''
state $\tr \mathcal S^{L/2}$ 
(one might be then tempted to call these spiky objects ``giant spinons'').
The ``antiferromagnetic'' state is the singlet
state of the $SO(6)$ spin-chain, and located at ``the farthest from BPS'' in the spin-chain spectrum.
These features can be
compared to that of magnons in the large spin sector (impurity
above BPS vacuum) corresponding to the transverse excitations of the
point-like string orbiting around a great circle of $S^{5}$\,.

\paragraph{}
It would be interesting to check the prediction (\ref{c}) directly
by using the conjectured AdS/CFT Bethe ansatz equation. In the
$SU(2)$ sector where the number of operators is finite, the nature
of the antiferromagnetic state is better understood
\cite{Rej:2005qt}, and the upper bound on the energy is known
\cite{Zarembo:2005ur} (see also \cite{Roiban:2006jt}). It is
proportional to $\sqrt{\lam}$\,, which is the same behavior as our
conjecture (\ref{c}). Recall that we argued the $SO(6)$ singlet
state was dual to a large winding string state with
zero-spins, (\ref{rational static prof}). If the prediction
(\ref{c}) is correct, then we should be able to reproduce it by
the $SO(6)$ Bethe ansatz equation approach. An approach similar to
\cite{Zarembo:2005ur} would be useful. In this case, the ``spiky
magnon'' part ``$Z^{K}\,\overline{Z}{}^{K'}\,W^{M}+\dots$'' could be
understood as (macroscopic number of) ``holes'' made in the
continuous mode numbers associated with the $SO(6)$ singlet Bethe
root configuration.\footnote{\, In the weak coupling regime, the
$SO(6)$ singlet Bethe root configuration and excitations above it
were studied in \cite{Minahan:2002ve, Engquist:2003rn,
Minahan:2004ds}.}
The $SO(6)$ singlet state was also studied in \cite{Rej:2007vm}, where an integral equation for the Bethe root density was derived.
It would be interesting to study it at strong coupling and compare it with our results.\footnote{\,
We thank M.~Staudacher for pointing this out to us.
}

Since the $\ts$ transformed string solutions discussed in this
paper are periodic classical solutions, one can define
corresponding action variables, namely the oscillation numbers. By
imposing the Bohr-Sommerfeld quantization condition, one obtains
integer valued action variables, which from lesson of the large
spin sector \cite{Chen:2006ge} we can again expect to correspond
to filling fractions defined for the $SO(6)$ spin-chain. It would
be interesting to understand this correspondence from the
finite-gap perspective along the lines of \cite{Dorey:2006zj,
Dorey:2006mx}.

It would be also interesting to compare the spectra of AdS/CFT
near the $SO(6)$ ``antiferromagnetic'' vacuum by an effective
sigma model approach (without any apparent use of integrability)
\cite{Kruczenski:2003gt}. In the $SU(2)$ case, a similar approach
was taken in \cite{Roiban:2006jt}, where a continuum limit of the
half-filled Hubbard chain was compared to an effective action for
``slow-moving'' strings with $J_{1}=J_{2}$\,. In our case, some
Hubbard-like model with $SO(6)$ symmetry would give clues.

We hope to revisit these issues in other publications in the near future.

\paragraph{Note added.}
After the submission of the first version of our paper to arXiv.org {\tt 0709.4033 [hep-th]} for publication, we learned that the paper {\tt 0709.4231 [hep-th]} \cite{Dimov:2007ey} appeared, in which single-spike strings are generalized to three-spin cases.
We thank N.~P.~Bobev and R.~C.~Rashkov for correspondence.

\subsubsection*{Acknowledgments}

We acknowledge useful discussions with N.~Dorey, Y.~Hatsuda,
Y.~Imamura, J.~Minahan, M.~Staudacher, A.A.~Tseytlin and K.~Zarembo. We thank
Y.~Imamura, J.~Minahan and A.A.~Tseytlin for reading the draft carefully and
giving us illuminating comments. KO is grateful to University of
Cambridge, Centre for Mathematical Sciences, for its warm
hospitality during the work was done. RS thanks the Yukawa
Institute for Theoretical Physics at Kyoto University. Discussions
during the YITP workshop YITP-W-07-05 on ``String Theory and
Quantum Field Theory'' were useful to complete this work. The work
of KO is supported in part by JSPS Research Fellowships for Young
Scientists. The work of BV was supported by EPSRC.

\appendix
\section*{Appendices}

\section{Helical Strings on \bmt{AdS_{3}\times S^{1}}}\label{app:AdS helicals}

This appendix is devoted to helical string solutions in the
$SL(2)$ sector. The construction almost parallels that in
\cite{Okamura:2006zv}, however, non-compactness of the AdS space
lead to new non-trivial features compared to the sphere case.

\subsection{Classical Strings on \bmt{AdS_{3}\times S^{1}} and Complex sinh-Gordon Model}

A string theory on $AdS_{3}\times S^{1}\subset AdS_{5}\times
S^{5}$ spacetime is described by an $O(2,2)\times O(2)$ sigma
model. Let us denote the coordinates of the embedding space as
$\eta_0$\,, $\eta_1$ (for $AdS_{3}$) and $\xi_{1}$ (for $S^{1}$)
and set the radii of $AdS_{3}$ and $S^{1}$ both to unity,
\begin{equation}
\vec \eta \, {}^* \cdot \vec \eta \equiv
- \abs{\eta_0}^2 + \abs{\eta_1}^2 = - 1\,, \qquad
\abs{\xi_{1}}^2 = 1\,.
\label{norms}
\end{equation}
In the standard polar coordinates, the embedding coordinates are expressed as
\begin{alignat}{3}
\eta_0 &= \cosh \rho \, e^{i \ssp t} \,, &\quad
\eta_1 &= \sinh \rho \,  e^{i \ssp \phi_1}\,, &\quad
\xi_{1} &= e^{i \ssp \varphi_1} \,,
\end{alignat}
and all the charges of the string states are defined as N\"other
charges associated with shifts of the angular variables. The
bosonic Polyakov action for the string on $AdS_{3}\times S^{1}$ is
given by
\begin{equation}
S = - \frac{\sqrt{\lambda}}{4\pi} \int d \sigma d \tau \cpare{
\gamma^{a b} \pare{ \partial_a \vec \eta \, {}^* \cdot \partial_b \vec \eta + \partial_a  \xi^* \cdot \partial_b \xi \,}
+ \widetilde\Lambda \Big( \vec \eta \, {}^* \cdot \vec \eta + 1 \Big)
+ \Lambda \Big( \xi_{1}^* \cdot  \xi_{1} - 1 \Big) }\,,
\label{action-AdS}
\end{equation}
and we take the same conformal gauge as in the \RS{3} case.
From the action (\ref{action-AdS}) we get the equations of motion
\begin{equation}
\partial_{a} \partial^{\ssp a} \vec \eta - (\partial_{a} \vec\eta \, {}^* \cdot \partial^{\ssp a} \vec \eta ) \, \vec \eta = 0\,, \qquad
\partial_{a} \partial^{\ssp a} \xi_{1} +(\partial_{a}  \xi_{1}^{*} \cdot \partial^{\ssp a}  \xi_{1}) \, \xi_{1} = 0 \,,
\label{str_eom}
\end{equation}
and Virasoro constraints
\begin{align}
0 &= {\cal T}_{\sigma \sigma} = {\cal T}_{\tau\tau} = \frac{\delta^{ab}}{2} \pare{ \partial_{a} \vec \eta \, {}^* \cdot \partial_{b} \vec \eta +
\partial_{a} \xi_{1}^* \cdot \partial_{b}  \xi_{1} }
\label{str_Vir1} \,, \\[2mm]
0 &= {\cal T}_{\tau\sigma} = {\cal T}_{\sigma\tau} = \Re \pare{ \partial_{\tau} \vec \eta \, {}^* \cdot \partial_{\sigma} \vec \eta +
\partial_{\tau}\xi_{1} \cdot \partial_{\sigma}\xi_{1}^{*} } \,.
\label{str_Vir2}
\end{align}

The PLR reduction procedure, which we made use of in obtaining the
$O(4)$ sigma model solutions from Complex sine-Gordon solution,
also works for the current case in much the same way. The $O(2,2)$
sigma model in conformal gauge is now related to what we call
Complex sinh-Gordon (CshG) model, which is defined by the
Lagrangian
\begin{equation}
{\cal L}_{\rm CshG} = \frac{\partial^{\ssp a} \psi^* \partial_a \psi}{1 + \psi^* \psi} + \psi^* \psi \,,
\label{CshG Lag}
\end{equation}
with $\psi = \psi (\tau, \sigma)$ being a complex field. It can be
viewed as a natural generalization of the well-known sinh-Gordon
model in the sense we describe below. By defining two real fields
$\alpha$ and $\beta$ of the CshG model through $\psi \equiv \sinh
\pare{\alpha/2} \exp (i \beta/2)$\,, the Lagrangian \eqref{CshG
Lag} is rewritten as
\begin{equation}
{\cal L}_{\rm CshG} = \frac14 \pare{\partial_a \alpha}^2 + \frac{\tanh^2 (\alpha/2)}{4} \pare{\partial_a \beta}^2 + \sinh^2 (\alpha/2)\,.
\label{CshG Lag2}
\end{equation}
The equations of motion that follow from the Lagrangian are
\begin{align}
&\partial^{\ssp a} \partial_a \psi - \psi^* \frac{\partial^{\ssp a} \psi \, \partial_a \psi}{1 + \psi^* \psi} - \psi \pare{1 + \psi^* \psi}=0\,,\label{CshG eom}\\[3mm]
&\quad \mbox{\em i.e.},\quad
\left\{
\begin{array}{l}
\ds \partial^{\ssp a} \partial_a \alpha - \frac{\sinh (\alpha/2)}{2 \cosh^3 (\alpha/2)} \pare{\partial_a \beta}^2 - \sinh \alpha = 0\,, \\[6mm]
\ds  \partial^{\ssp a} \partial_a \beta + \frac{2 \, \partial_a \alpha \, \partial^{\ssp a} \beta}{\sinh \alpha} = 0\,.
\end{array}
\right.\label{CshG eq}
\end{align}
We refer to the coupled equations \eqref{CshG eq} as Complex
sinh-Gordon (CshG) equations. If $\beta$ is a constant field, the
first equation in (\ref{CshG eq}) reduces to
\begin{equation}
\partial_a \partial^{\ssp a} \alpha - \sinh \alpha = 0\,.
\end{equation}
which is the ordinary sinh-Gordon equation. As readers familiar
with the PLR reduction can easily imagine, it is this field
$\alpha$ that gets into a self-consistent potential in the
Schr\"{o}dinger equation this time. Namely, we can write the
string equations of motion given in \eqref{str_eom} as
\begin{equation}
\partial_{a} \partial^{\ssp a} \vec \eta - (\cosh \alpha) \, \vec \eta = 0\,, \qquad
\cosh \alpha \equiv \partial_{a} \vec\eta \, {}^* \cdot \partial^{\ssp a} \vec \eta\,,
\label{reduced_eom}
\end{equation}
with the same field $\alpha$ we introduced as the real part of the
CshG field $\psi$\,. What this means is that if $\{\vec \eta\,,
\xi\}$ is a consistent string solution which satisfies Virasoro
conditions \eqref{str_Vir1} and \eqref{str_Vir2}, then $\psi =
\sinh \pare{\alpha/2} \exp (i \beta/2)$ defined via
\eqref{reduced_eom} and \eqref{AdS-PLR beta} solves the CshG
equations.

The derivation of this fact parallels the usual PLR reduction
procedure. Let us define worldsheet light-cone coordinates as
$\sigma^{\pm}=\tau\pm\sigma$\,, and the embedding coordinates as
$\eta_0 = Y_0 + i Y_5$ and $\eta_1 = Y_1 + i Y_2$\,. Then consider
the equations of motion of the $O(2,2)$ nonlinear sigma model
through the constraints
\begin{equation}
\vec Y \cdot \vec Y = -1\,, \quad (\partial_+ \vec Y )^2 = -1\,,\quad
(\partial_- \vec Y )^2 = -1\,,\quad \partial_+ \vec Y \cdot \partial_- \vec Y \equiv - \cosh \alpha\,,
\end{equation}
where $\vec Y \cdot \vec Y \equiv (\vec Y)^2 \equiv - (Y_0)^2 + (Y_1)^2 + (Y_2)^2 - (Y_5)^2$\,.
A basis of $O(2,2)$-covariant vectors can be given by $Y_i$\,, $\partial_+ Y_i $\,, $\partial_- Y_i$ and $K_i \equiv \epsilon_{ijkl} Y^j \partial_+ Y^k \partial_- Y^l$\,.
By defining a pair of scalar functions $u$ and $v$ as
\begin{equation}
u \equiv \frac{\vec K \cdot \partial_+ ^{\ssp 2} \vec Y}{\sinh
\alpha} \,, \qquad v \equiv \frac{\vec K \cdot \partial_- ^{\ssp
2} \vec Y}{\sinh \alpha} \,,
\end{equation}
the equations of motion of the $O(2,2)$ sigma model are recast in
the form
\begin{equation}
\partial_- \partial_+ \alpha + \sinh \alpha + \frac{uv}{\sinh \alpha} = 0\,, \qquad
\partial_{-} u = \frac{v \, \partial_{+} \alpha}{\sinh \alpha} \,, \qquad
\partial_{+} v = \frac{u \, \partial_{-} \alpha}{\sinh \alpha} \,.
\label{Pohlmeyer_eom1}
\end{equation}
One can easily confirm that this set of equations is equivalent to
the pair of equations (\ref{CshG eq}) of CshG theory, under the
identifications
\begin{equation}
u = (\partial_+ \beta) \, \tanh \frac{\alpha}{2} \,,\qquad
v = - (\partial_- \beta) \, \tanh \frac{\alpha}{2} \,.
\label{AdS-PLR beta}
\end{equation}

\paragraph{}
Thus there is a (classical) equivalence between the $O(2,2)$ sigma
model $\leftrightarrow$ CshG as in the $O(4)\leftrightarrow
\mbox{CsG}$ case. Making use of the equivalence, one can construct
classical string solutions on $AdS_{3}\times S^{1}$ by the
following recipe\,:
\begin{enumerate}
\item Find a solution $\psi$ of CshG equation (\ref{CshG eom}).
\item Identify $\cosh\alpha\eq \partial_{a} \vec\eta \, {}^* \cdot
\partial^{\ssp a} \vec \eta$\,, where $\alpha$ appears in the real
part of the solution $\psi$\,, and $\eta$ are the embedding
coordinates of the corresponding string solution in $AdS_{3}$\,.
\item Solve the ``Schr\"{o}dinger equation'' (\ref{reduced_eom})
together with the Virasoro constraints (\ref{str_Vir1}) and
(\ref{str_Vir2}), under appropriate boundary conditions. \item
Resulting set of $\vec\eta$ (``wavefunction'') and $\xi_{1}$ gives
the corresponding string profile in $AdS_{3}\times S^{1}$\,.
\end{enumerate}

Let us start with step 1. From the similarities between the CshG
equation and the CsG equation, it is easy to find helical-wave
solutions of the CshG equation. Here we give two such solutions
that will be important later. The first one is given by
\begin{equation}
\psi_{\rm cd}=kc\frac{\cn (c x_v)}{\dn (c x_v)}\exp \Big( i \sqrt{(1+c^2)(1+k^2 c^2)} \; t_v \Big)\,,
\label{CshG helical cd}
\end{equation}
and the second one is
\begin{equation}
\psi_{\rm ds}=c\frac{\dn (c x_v)}{\sn (c x_v)}\exp  \Big( i \sqrt{(1-k^2c^2)(1+c^2-k^2c^2)} \; t_v \Big)\,.
\label{CshG helical ds}
\end{equation}
By substituting the solution (\ref{CshG helical ds}) into the string equations of motion \eqref{reduced_eom}, we obtain
\begin{equation}
\cpare{- \partial_T ^2 + \partial_X ^2 - k^2 \pare{\frac{2}{k^2 \sn^2 (X,k)} - 1} } \, \vec \eta = U \vec \eta \,,
\label{GAL eom}
\end{equation}
under the identification of $(\mu \ssp \tau, \mu \ssp \sigma)
\equiv (c \ssp t, c \ssp x)$\,. The ``eigenenergy'' $U$ can be
treated as a free parameter as was the case in
\cite{Okamura:2006zv}. Different choices of helical-waves of CshG
equation simply correspond to taking different ranges of $U$\,.

\paragraph{}
We are now at the stage of constructing the corresponding string
solution by following the steps 2\,\--\,4 listed before. However,
we do not need to do this literally. Since the metrics of \AdSS{3}
\begin{equation}
ds_{AdS_{3}\times S^{1}}^2 = - \cosh^2 \! \tilde \rho \, d {\tilde t\,}{}^2 + d{\tilde \rho}^2 + \sinh^2 \! \tilde \rho \, d {\tilde \phi_1}^2 + d {\tilde \varphi_1}^2 \,,
\end{equation}
and of \RS{3}
\begin{equation}
ds_{\mathbb R\times S^{3}}^2 = - d t^2 + d\gamma^2 + \cos^2 \! \gamma \, d \varphi_1^2 + \sin^2 \! \gamma \, d \varphi_2^2 \,,
\end{equation}
are related by analytic continuation
\begin{equation}
\tilde\rho  \leftrightarrow i\gamma ,\quad \tilde t \leftrightarrow \varphi _1 ,\quad \tilde \phi _1  \leftrightarrow \varphi _2 ,\quad \tilde \varphi _1  \leftrightarrow t \quad \Longrightarrow \quad ds_{AdS_{3}\times S^{1}}^2 \leftrightarrow - ds_{\mathbb R\times S^{3}}^2 \,,
\label{an_cont}
\end{equation}
string solutions on both manifolds are related by a sort of
analytic continuation of global coordinates. Therefore, the
simplest way to obtain helical string solutions on \AdSS{3} is to
perform analytic continuation of helical string solutions on
\RS{3}, as will be done in the following sections. Large parts of
the calculation parallel the $\mathbb R \times S^{3}$ case. The
most significant difference lies in the constraints imposed on the
solution of the equations of motion, such as the periodicity
conditions.

\subsection{Helical Strings on \bmt{AdS_{3}\times S^{1}} with Two Spins}

In this section, we consider the analytic continuation of helical
strings on \RS{3} to those on \AdSS{3}. Among various possible
solutions, we will concentrate on two particular examples that
have clear connections with known string solutions of interest to
us. The first example, called type $(iii)$ helical string, is a
helical generalization of the folded string solution on \AdSS{3}
\cite{Frolov:2002av}. The second one, called type $(iv)$,
reproduces the $SL(2)$ ``giant magnon" solution
\cite{Minahan:2006bd, Ryang:2006yq} in the infinite-spin limit.

\subsubsection{Type \bmt{(iii)} Helical Strings}

In \cite{Beisert:2003ea}, it was pointed out that $(S, J)$ folded
strings can be obtained from $(J_1, J_2)$ folded strings by
analytic continuation of the elliptic modulus squared, from $k^2
\ge 0$ to $k^2 \le 0$\,. Here we apply the same analytic
continuation to type $(i)$ helical strings to obtain solutions on
\AdSS{3}, which we call type $(iii)$ strings. For notational
simplicity, it is useful to introduce a new moduli parameter $q$
through the relation
\begin{equation}
k \equiv \frac{i q}{q'} \eq \frac{i q}{\sqrt{1 - q^2}} \,.
\label{moduli kq}
\end{equation}
If $k$ is located on the upper half of the imaginary axis, {\em
i.e.}, $k=i\kappa$ with $0\leq \kappa$\,, then $q$ is a real
parameter in the interval $[0,1]$\,.

\begin{figure}[htbp]
\begin{center}
\vspace{0.5cm}
\includegraphics[scale=0.9]{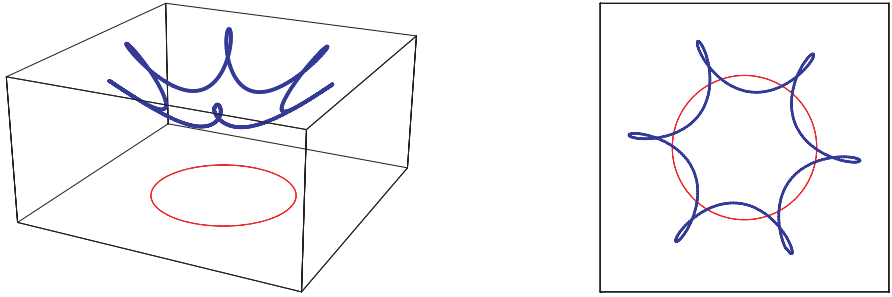}
\vspace{0.5cm}
\caption{\small Type $(iii)$ helical string ($q=0.700$\,, $U=12.0$\,, $\tilde \om_{0}=-0.505$\,, $\tilde\om_{1}=0.776$\,, $n=6$), projected onto $AdS_{2}$ spanned by $(\Re\!\eta_{1}, \Im\!\eta_{1}, |\eta_{0}|)$\,.
The circle represents a unit circle $|\eta_{1}|=1$ at $\eta_{0}=0$\,.}
\label{fig:III-gen}
\end{center}
\end{figure}

As shown in Appendix \ref{app:formula}, the transformation
\eqref{moduli kq} can be regarded as a $\mathsf T$-transformation
of the modulus $\tau$. Hence, by performing a $\mathsf
T$-transformation on the profile of type $(i)$ helical strings
\eqref{zf0}-\eqref{zf2}, we obtain type $(iii)$ string solutions:
\begin{align}
\eta_0 &= \frac{C}{\sqrt{q q'}} \, \frac{\Theta_3 (0) \, \Theta_0 (\tilde X - \tomm{0})}{\Theta_2 (\tomm{0}) \, \Theta_3 (\tilde X)} \,
\exp \Big( Z_2 (\tomm{0}) \tilde X + i \tilde u_0 \ssp \tilde T \Big) \,,  \label{e0_two} \\[2mm]
\eta_1 &= \frac{C}{\sqrt{q q'}} \, \frac{\Theta_3 (0) \, \Theta_1 (\tilde X - \tomm{1})}{\Theta_3 (\tomm{1}) \, \Theta_3 (\tilde X)} \,
\exp \Big( Z_3 (\tomm{1}) \tilde X + i \tilde u_1 \ssp \tilde T \Big) \,,  \label{e1_two} \\[2mm]
\xi_1 &= \exp \pare{ i \ssp \tilde a \ssp \tilde T + i \ssp \tilde b \ssp \tilde X } \,,  \label{x1_two}
\end{align}
where we rescaled various parameters as
\begin{equation}
\tilde X = X/q' \,,\quad \tilde T = T/q' \,,\quad \tilde \omega_j = \omega_j / q' \,, \quad \tilde a = a \ssp q'\,, \quad \tilde b = b \ssp q' \,, \quad \tilde u_j = u_j \ssp q' \,.   \label{var tildes}
\end{equation}
We choose the constant $C$ so that they satisfy $\left| \eta_0
\right|^2 - \left| \eta_1 \right|^2 = 1$\,. One such possibility
is to choose\footnote{\, In contrast to the \RS{3} case, the RHS
of (\ref{cd_two}) is not always real for arbitrary real values of
$\tilde \omega_0$ and $\tilde \omega_1$\,. If $C^2 < 0$\,, we have
to interchange $\eta_0$ and $\eta_1$ to obtain a solution properly
normalized on \AdS{3}\,. }
\begin{equation}
C = \pare{\frac{1}{q^2 \cn^2 (\tomm{0})}  + \frac{\sn^2 (\tomm{1})}{\dn^2 (\tomm{1})} }^{-1/2} \,.  \label{cd_two}
\end{equation}
With the help of various formulae on elliptic functions, one can check that $\vec \eta$ in (\ref{e0_two}), (\ref{e1_two}) certainly solves the string equations of motion as
\begin{equation}
\cpare{- \partial_{\tilde T} ^2 + \partial_{\tilde X} ^2 + q^2 \pare{2 (1-q^2) \, \frac{\sn^2}{\dn^2} (\tilde X,q) - 1 }} \, \vec \eta = \tilde U \vec \eta \,,
\label{str eom_two}
\end{equation}
if the parameters are related as
\begin{equation}
\tilde u_0^2 = \tilde U - (1-q^2) \frac{\sn^2 (\tomm{0})}{\cn^2 (\tomm{0})} \,,\qquad
\tilde u_1^2 = \tilde U + \frac{1 - q^2}{\dn^2 (\tomm{1})} \,.
\end{equation}
As is clear from \eqref{str eom_two}, the type $(iii)$ solution is
related to the helical-wave solution of the CshG equation given in
\eqref{CshG helical cd}. The Virasoro constraints \eqref{str_Vir1}
and \eqref{str_Vir2} impose constraints on $\tilde a$ and $\tilde
b$ in (\ref{x1_two})\,:\footnote{\, Note that the Virasoro
constraints require neither $a \ge b$ nor $a \le b$\,. This means
that both $\xi_1 = \exp \big(i \tilde a_0 \tilde T + i \tilde b_0
\tilde X\big)$ and $\exp\big(i \tilde b_0 \tilde T + i \tilde a_0
\tilde X\big)$ are consistent string solutions. It can be viewed
as the $\ts$ transformation applied only to the $S^{1}\subset
S^{5}$ part while leaving the \AdS{3} part intact.}
\begin{alignat}{2}
&\tilde a^2 + \tilde b^2 & &= - q^2 - \tilde U - \frac{2 \ssp (1 - q^2)}{\cn^2 (\iomm{0})} + 2 \ssp \tilde u_1^2 \,,
\label{ab1_two}  \\
&\quad \tilde a \ssp \tilde b & &= i \, C ^2 \pare{\frac{\tilde u_0}{q^2} \, \frac{\sn (\iomm{0}) \dn (\iomm{0})}{\cn^3 (\iomm{0})} + \tilde u_1 \, \frac{\sn (\iomm{1}) \cn (\iomm{1})}{\dn^3 (\iomm{1})}}
\label{ab2_two}\,.
\end{alignat}
The reality of $\tilde a$ and $\tilde b$ must also hold.

Since we are interested in closed string solutions, we should
impose periodic boundary conditions. Let us define the period in
the $\sigma$ direction by
\begin{equation}
\Delta \sigma = \frac{2 \ssp \eK (k) \sqrt{1 - v^2}}{\mu}
= \frac{2 \ssp q' \, \eK (q) \sqrt{1 - v^2}}{\mu} \equiv 2 l \equiv \frac{2 \pi}{n}\,,
\label{one-hop def}
\end{equation}
which is equivalent to $\Delta \tilde X = 2 \ssp \eK (q)$ and $\Delta \tilde T = - 2 \ssp v \ssp \eK (q)$.
The closedness conditions for the AdS variables are written as
\begin{alignat}{2}
\Delta t &= 2 \eK (q) \bpare{ -i Z_2 (\tomm{0}) - v \ssp \tilde u_0 } + 2 \ssp n'_{\rm time} \ssp \pi \equiv \frac{2 \pi N_t}{n}\,,
\label{Dt_ads_dy} \\[2mm]
\Delta \phi_1 &= 2 \eK (q) \bpare{ -i Z_3 (\tomm{1}) - v \ssp \tilde u_1 } + \pare{2 \ssp n'_1 + 1} \pi \equiv \frac{2 \pi N_{\phi_{1}}}{n} \,.
\label{Dphi_ads_dy}
\end{alignat}
And from the periodicity in $\varphi_1$ direction, we have
\begin{equation}
N_{\varphi_{1}} = \mu \, \frac{\tilde b - v \ssp \tilde a}{\sqrt{1 - v^2}} \in \bb{Z} \,.
\label{Dvarphi_ads_dy}
\end{equation}

We must further require the timelike winding $N_t$ to be zero. Just as in the \RS{3} case, one can adjust the value of $v$ to fulfill this requirement.\footnote{\,Note in $\mathbb R\times S^{3}$ case, the vanishing-$N_t$ condition was trivially solved by $v=b/a$\,.} 
The integer $n'_{\rm time}$ is evaluated as
\begin{equation}
2 \ssp n'_{\rm time} \ssp \pi = \frac{1}{2 \ssp i} \int_{-\eK}^{\eK} d\tilde X \; \frac{\partial}{\partial \tilde X} \cpare{ \log \pare{\frac{\Theta_0 (\tilde X - \tomm{0})}{\Theta_0 (\tilde X + \tomm{0})}} } \,.
\end{equation}
Then, by solving the equation $N_t=0$\,, one finds an appropriate
value of $v=v_t$. The absolute value of the worldsheet boost
parameter $v_t$ may possibly exceed one (the speed of light). In
such cases, we have to perform the 2D transformation $\tau
\leftrightarrow \sigma$ on the AdS space to get $v_t \mapsto
-1/v_t$\,.

\bigskip
As usual, conserved charges are defined by
\begin{alignat}{3}
E &\equiv \frac{\sqrt{\lambda}}{\pi} \, {\cal E} & &= \frac{n \sqrt{\lambda}}{2 \pi} \int_{-l}^{\ssp l} d \sigma \,
\Im\!\pare{\eta_0^{*} \, \partial_{\tau} \eta_0} \,,  \label{charge_def E}\\
S &\equiv \frac{\sqrt{\lambda}}{\pi} \, {\cal S} & &= \frac{n \sqrt{\lambda}}{2 \pi} \int_{-l}^{\ssp l} d \sigma
\Im\!\pare{\eta_{1}^{*} \, \partial_{\tau} \eta_{1}} \,,  \label{charge_def S}\\
J &\equiv \frac{\sqrt{\lambda}}{\pi} \, {\cal J} & &= \frac{n \sqrt{\lambda}}{2 \pi} \int_{-l}^{\ssp l} d \sigma
\Im\!\pare{\xi_{1}^{*} \, \partial_{\tau} \xi_{1}} \,.  \label{charge_def J}
\end{alignat}
which are evaluated as, for the current type $(iii)$ case,
\begin{align}
{\cal E} &= \frac{n \ssp C^2 \, \tilde u_0}{q^2 (1-q^2)} \cpare{ \eE
+ (1-q^2) \bpare{ \frac{\sn^2 (\tomm{0})}{\cn^2 (\tomm{0})} - \frac{iv}{\tilde u_0} \, \frac{\sn (\tomm{0}) \dn (\tomm{0})}{\cn^3 (\tomm{0})} } \eK } \,,  \label{chE two}  \\[2mm]
{\cal S} &= \frac{n \ssp C^2 \, \tilde u_1}{q^2 (1-q^2)} \cpare{ \eE
- (1-q^2) \bpare{ \frac{1}{\dn^2 (\tomm{1})} - \frac{iv \ssp q^2}{\tilde u_1} \, \frac{\sn (\tomm{1}) \cn (\tomm{1})}{\dn^3 (\tomm{1})} } \eK } \,,  \label{chS two}  \\[2mm]
{\cal J} &= n \pare{\tilde a - v \ssp \tilde b} \eK \,.
\label{chJ two}
\end{align}

\subsubsection*{}
It is interesting to see some of the limiting behaviors of this
type $(iii)$ helical string in detail.\footnote{\, It seems the
original ``spiky string'' solution of \cite{Kruczenski:2004wg} is
also contained in the type $(iii)$ class, although we have not been able to reproduce it analytically. 
}

\subsubsection*{$\bullet$ \bmt{\tilde \om_{1,2}\to 0} limit\,:~Folded strings on \bmt{AdS_{3}\times S^{1}}}

In the $\tilde \om_{1,2}\to 0$ the timelike winding condition \eqref{Dt_ads_dy} requires $v = 0$\,, so the boosted worldsheet coordinates $(\tilde T, \tilde X)$ become
\begin{equation}
(\tilde T, \tilde X) \to \pare{\frac{\mu \tau}{q'}\,, \frac{\mu \sigma}{q'}} \equiv
(\tilde \mu \tau, \tilde \mu \sigma) \equiv (\tilde \tau, \tilde \sigma) \,.
\label{tilde ts}
\end{equation}
The periodicity condition \eqref{one-hop def} allows $\tilde \mu$
to take only a discrete set of values.

\begin{figure}[htbp]
\begin{center}
\vspace{0.5cm}
\includegraphics[scale=0.9]{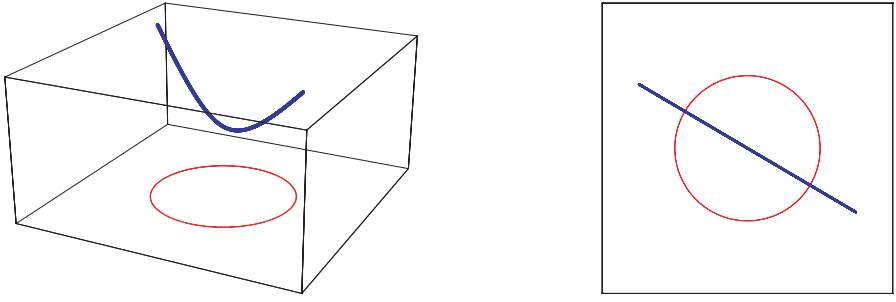}
\vspace{0.5cm}
\caption{\small $\tilde \om_{1,2}\to 0$ limit of type $(iii)$ helical string becomes a folded string studied in \cite{Frolov:2002av}.}
\label{fig:III-fold}
\end{center}
\end{figure}

The profile of type $(iii)$ strings now reduces to
\begin{equation}
\eta_0 = \frac{1}{\dn(\tilde \sigma, q)} \; e^{i \tilde u_0 \tilde \tau} \,, \qquad
\eta_1 = \frac{q \sn(\tilde \sigma, q)}{\dn(\tilde \sigma, q)} \; e^{i \tilde u_1 \tilde \tau} \,, \qquad
\xi_1 = \exp \pare{ i \sqrt{\tilde U - q^2} \; \tilde \tau}\,,
\label{stat_i}
\end{equation}
where $\tilde u_0^2 = \tilde U$ and $\tilde u_1^2= \tilde U + 1 - q^2$\,.
This solution is equivalent to $\mathsf T$-transformation
of $(J_1, J_2)$ folded strings of \cite{Frolov:2003xy}, namely, $(S, J)$ folded strings.\footnote{\,
Note the set, $\eta_{0,1}=\mbox{the same as (\ref{stat_i})}$ and $\xi_{1}=\exp[ i \sqrt{\tilde U - q^2} \; \tilde \sigma]$\,, also gives a solution.}
The conserved charges of \eqref{stat_i} are computed as
\begin{equation}
{\cal E} = \frac{n \tilde u_0}{1-q^2} \, \eE (q) \,, \quad
{\cal S} = \frac{n \tilde u_1}{1-q^2} \Big( \eE (q) - (1-q^2) \eK (q) \Big) \,,\quad
{\cal J} = n \sqrt{\tilde U - q^2} \; \eK (q) \,.
\end{equation}
Rewriting these expressions in terms of the original imaginary
modulus $k$\,, we find the following relations among conserved
charges\,:
\begin{equation}
\left( {\frac{\cal J}{{\eK (k)}}} \right)^2  - \left( {\frac{\cal E}{{\eE (k)}}} \right)^2  = n^2 k^2 \,,\qquad
\left( {\frac{\cal S}{{\eK (k) - \eE (k)}}} \right)^2  - \left( {\frac{\cal J}{{\eK (k)}}} \right)^2  = n^2 (1 - k^2) \,,
\label{FT relation}
\end{equation}
as obtained in \cite{Beisert:2003ea}.

\subsubsection*{$\bullet$ \bmt{q\to 1} limit\,:~ Logarithmic behavior}

Another interesting limit is to send the elliptic modulus $q$ to unity.
In this limit, the spikes of the type $(iii)$ string attach to the AdS boundary, and the energy $E$ and AdS spin $S$ become divergent.
Again, the condition of vanishing timelike winding is fulfilled by $v = 0$, and the periodicity condition \eqref{one-hop def} implies that $\tilde \mu$ given in \eqref{tilde ts} goes to infinity.
The profile becomes
\begin{equation}
\eta_0 = C \cosh(\tilde \sigma - \tomm{0}) \; e^{i \tilde u_0 \tilde \tau} \,, \quad
\eta_1 = C \sinh(\tilde \sigma - \tomm{1}) \; e^{i \tilde u_1 \tilde \tau} \,, \quad
\xi_1 = \exp \pare{ i \ssp \tilde a \ssp \tilde \tau + i \ssp \tilde b \ssp \tilde \sigma } \,,
\end{equation}
where
\begin{equation}
C = \pare{\cos^2 \tilde \omega_1 - \sin^2 \tilde \omega_0}^{-1/2}\,, \quad
\tilde u_0^2 = \tilde u_1^2 = \tilde U \,.
\label{inf iii consts}
\end{equation}
The constants $\tilde a$ and $\tilde b$ satisfy the constraints
\begin{equation}
\tilde a^2 + \tilde b^2 = - 1 + \tilde U\,, \qquad
\tilde a \ssp \tilde b = C^2 \pare{ \tilde u_0 \ssp \sin \tilde\omega_0 \cos \tilde\omega_0 + \tilde u_1 \ssp \sin \tilde\omega_1 \cos \tilde\omega_1 } \,.
\end{equation}
The conserved charges are computed as
\begin{equation}
{\cal E} = n \ssp C^2 \ssp \tilde u_0 \Big( \Lambda - \sin^2 \tilde \omega_0 \, \eK (1) \Big) \,, \quad
{\cal S} = n \ssp C^2 \ssp \tilde u_1 \Big( \Lambda - \cos^2 \tilde \omega_1 \, \eK (1) \Big) \,, \quad
{\cal J} = n \ssp \tilde a \, \eK (1)\,,
\end{equation}
where we defined a cut-off $\Lambda \eq 1/(1 - q^2)$\,.

\paragraph{}
Let us pay special attention to the $\tilde u_0 = \tilde u_1 =
\sqrt{\tilde U}$ case. For this case the energy-spin relation
reads
\begin{equation}
{\cal E} - {\cal S} = n \sqrt{\tilde U} \; \eK (1)\,.
\label{inf iii E-S}
\end{equation}
Obviously the RHS is divergent, and careful examination reveals it
is logarithmic in ${\cal S}$\,. This can be seen by first
noticing, on one hand, that the complete elliptic integral of the
first kind $\eK (q) \equiv \eK (e^{-r})$ has asymptotic behavior
\begin{equation}
\eK (e^{-r}) = - \frac12 \ln \pare{\frac{r}{8}} + {\cal O} (r \ln r) \,,
\end{equation}
while on the other, the degree of divergence for $\Lambda$ is
\begin{equation}
\Lambda = \frac{1}{1-q^2} = \frac{1}{1 - e^{-2r}} \sim \frac{1}{2r}\,, \qquad
({\rm as}\ r \to 0)\,.
\end{equation}
Since the most divergent part of ${\cal S}$ is governed by
$\Lambda$ rather than $\eK(1)$\,, it follows that
\begin{equation}
\eK (e^{-r}) \sim \eK (1 - r) \sim - \frac12 \, \ln \pare{\frac{n \ssp C^2 \ssp \tilde u_1}{16 \ssp {\cal S}}} \,, \qquad (\mbox{as}~ r \to 0)\,,
\label{eK lnS}
\end{equation}
at the leading order.
Then it follows that
\begin{equation}
{\cal E} - {\cal S} \sim - \frac{n \sqrt{\tilde U}}{2} \, \ln \pare{\frac{16 \cal S}{n \ssp C^2 \ssp \tilde u_1}} \,, \qquad (\mbox{as}~ r \to 0)
\label{e-s-lns}
\end{equation}
as promised.

Let us consider the particular case $\tilde U = 1$\,, which is
equivalent to $\tilde{a}=\tilde{b}=0$ and $\tilde{\omega}_0 =
-\tilde{\omega}_1$\,. The above dispersion relation
\eqref{e-s-lns} now reduces to
\begin{equation}
E - S \sim \frac{n \sqrt{\lambda}}{2\pi} \; \ln S \,,
\label{log div}
\end{equation}
omitting the finite part. This result was first obtained in
\cite{Gubser:2002tv} for the $n=2$ case, and generalised to
generic $n$ case in \cite{Kruczenski:2004wg}.

\paragraph{}
One can also reproduce the double logarithm behavior of \cite{Frolov:2002av} (see also \cite{Beisert:2003ea, Belitsky:2006en, Frolov:2006qe, Casteill:2007ct}).
To see this, let us set $\tilde b=0$ and $\tilde a=\sqrt{\tilde U-1}$\,, and rewrite the relation \eqref{inf iii E-S} as
\begin{equation}
{\cal E} - {\cal S} = \sqrt{ \mathstrut {\cal J}^2 + n^2 \; \eK (1)^2} \sim \kko{\mathstrut {\cal J}^2 + \frac{n^2}{4} \ln^2 \pare{\frac{2 {\cal S}}{n \ssp C^2 \ssp \sqrt{\tilde U} }}}^{1/2} \,.
\label{inf iii E-S 2}
\end{equation}
There are two limits of special interest. The ``slow long string"
limit of \cite{Frolov:2006qe}, is reached by $\sqrt{U} \ll
\lambda$\,, so that in the strong coupling regime $\lambda \gg 1$
the RHS of (\ref{inf iii E-S 2}) becomes
\begin{equation}
{\cal E} - {\cal S} \sim \sqrt{\mathstrut {\cal J}^2 + \frac{n^2}{4} \ln^2 {\cal S} } \,.
\label{inf iii slowlong}
\end{equation}
Similarly, the ``fast long string" of \cite{Frolov:2006qe} is obtained by taking $\sqrt{U} \sim \lambda \gg 1$\,, resulting in
\begin{equation}
{\cal E} - {\cal S} \sim \kko{\mathstrut {\cal J}^2 + \frac{n^2}{4} \ko{\ln \pare{\frac{\cal S}{\cal J}} + \ln \pare{\ln r}}^2 }^{1/2} \sim \sqrt{\mathstrut {\cal J}^2 + \frac{n^2}{4} \ln^2 \pare{\frac{\cal S}{\cal J}} } \,,
\label{inf iii fastlong}
\end{equation}
where we neglected a term $\ln \pare{\ln r}$ which is relatively less divergent in the limit $r \to 0$\,.

\subsubsection{Type \bmt{(iv)} Helical Strings}

Let us finally present another AdS helical solution which
incorporates the $SL(2)$ ``(dyonic) giant magnon'' of \cite{Minahan:2006bd,
Ryang:2006yq}. This solution, which we call the type $(iv)$
string, is obtained by applying a shift $X \to X + i \eK' (k)$ to
the type $(i)$ helical string. Its profile is given by
\begin{align}
\eta_0 &= \frac{C}{\sqrt{k}} \, \frac{\Theta _0 (0) \, \Theta _0 (X  - \iomm{0})}{\Theta_0 (\iomm{0}) \, \Theta _1 (X)} \,
\exp \Big( Z_0 (\iomm{0}) X + i u_0 T \Big) \,,
\label{e0d_two} \\[2mm]
\eta_1 &= \frac{C}{\sqrt{k}} \, \frac{\Theta _0 (0) \, \Theta _3 (X  - \iomm{1})}{\Theta_2 (\iomm{1}) \, \Theta _1 (X)} \,
\exp \Big( Z_3 (\iomm{1}) X + i u_1 T \Big) \,,
\label{e1d_two} \\[2mm]
\xi_1 &= \exp \pare{ i a T + i b X } \,.
\label{xi1d_two}
\end{align}
We omit displaying all the constraints among the parameters (they
can be obtained in a similar manner as in the type $(i)$ case).
The type $(iv)$ solution corresponds to the helical-wave solution
given in \eqref{CshG helical ds}, and satisfy the string equations
of motion of the form \eqref{GAL eom}. \footnote{\, This can be
easily checked by using a relation $1/k^2 \sn^2 (x,k) = \sn^2 \ko{
x +i\eK'(k), k }$\,. }

\subsubsection*{$\bullet$ \bmt{k\to 1} limit\,:~ \bmt{SL(2)} ``dyonic giant magnon"}

The $SL(2)$ ``dyonic giant magnon'' is reproduced in the limit $k\to 1$\,, as
\begin{equation}
\eta_{0}=\f{\cosh(X-\iomm{0})}{\sinh X}\; e^{i(\tan\om_{0})X+i u_{0}T}\,,\quad
\eta_{1}=\f{\cos\om_{0}}{\sinh X}\; e^{i u_{1}T}\,,\quad
\xi_{1}=e^{\hat a T + i\hat b X}\,,
\label{dual-GM}
\end{equation}
where
\begin{equation}
u_0^{2} = u_1^{2} + \frac{1}{\cos^{2} \omega_{0}} \,, \qquad
(\hat a, \hat b)=(u_{1},\tan\om_{0}) \ \ {\rm or}\ \ (\tan\om_{0}, u_{1})\,.
\end{equation}
Due to the non-compactness of AdS space, the conserved charges are
divergent. This is an UV divergence, and we regularise it by the following prescription.
First change the integration range for the charges (see (\ref{charge_def
E}) - (\ref{charge_def J})) from $\int_{0}^{2l}d\sig$ to
$\int_{\ep}^{2l-\ep}d\sig$\,, with $\epsilon>0$\,, to obtain
\begin{align}
\mathcal E &= u_{0}\cos^{2}\om_{0} \pare{\ep^{-1}-1} +\eK(1)(u_{0}-v\tan\om_{0})\,,\\
\mathcal S &= u_{1}\cos^{2}\om_{0} \pare{\ep^{-1}-1}\,,\\
{\mathcal J}&=\eK(1)(u_{0}-v\tan\om_{0})\,,
\end{align}
then drop the terms proportional to $\ep^{-1}$ by hand. This prescription yields
a regularised energy and an $S^{5}$ spin which are still IR divergent
due to the non-compactness of the worldsheet. However, their
difference becomes finite, leading to the energy-spin relation
\begin{equation}
(\mathcal E-\mathcal J)_{\rm reg}=-\sqrt{(\mathcal S)_{\rm reg}^{2}+\cos^{2}\om_{0}}\,.
\end{equation}
Note that in view of the AdS/CFT correspondence, ${\cal E} - {\cal J}$ must be positive, which in turn implies $({\cal E} - {\cal J})_{\rm reg}$ is negative.

\begin{figure}[htbp]
\begin{center}
\vspace{0.5cm}
\includegraphics[scale=0.9]{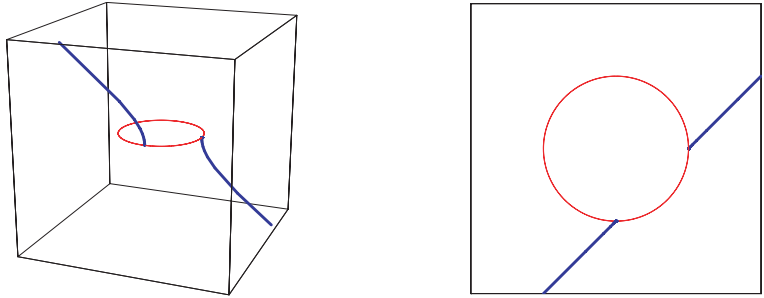}
\vspace{0.5cm}
\caption{\small $k\to 1$ limit of type $(iv)$ helical string ($\om_{0}=0.785$\,, $u_{0}=1.41$\,, $u_{1}=0$)\,: ``giant magnon'' solution in AdS space.}
\label{fig:dual-GM}
\end{center}
\end{figure}

Let us take $v = \tan \omega_0/u_0$ in (\ref{dual-GM}), and
consider a rotating frame ${\eta}_{0}^{\rm new} = e^{- i \tilde
\tau} \eta_0 \equiv \tilde Y_0 + i \tilde Y_5$\,. We then find
$\tilde Y_5=-i\sin\om_{0}$ is independent of $\tilde \tau$ and
$\tilde\sigma$\,, showing that the ``shadow'' of the $SL(2)$
``dyonic giant magnon'' projected onto the $\tilde Y_0$-$\tilde Y_5$ plane is just
given by two semi-infinite straight lines on the same line.
Namely, the shadow is obtained by removing a finite segment from
an infinitely long line, where the two endpoints of the segment
are on the unit circle $|\eta_{0}|=1$ with angular difference
$\Delta t = \pi - 2\omega_0$\,. Figure \ref{fig:dual-GM} shows the
snapshot of the $SL(2)$ ``dyonic giant magnon'', projected onto the plane
spanned by $(\Re\!\eta_{0}, \Im\!\eta_{0}, |\eta_{1}|)$\,.

It is interesting to compare this situation with the usual giant magnon on $\mathbb R\times S^3$\,.
In the sphere case, the ``shadow'' of the giant magnon is just a straight line segment connecting two endpoints on the equatorial circle $|\xi_{1}|=1$\,.
So the ``shadows'' of $SU(2)$ and $SL(2)$ giant magnons are just complementary.
Using this picture of ``shadows on the LLM plane'', one can further discuss the ``scattering'' of two $SL(2)$ ``(dyonic) giant magnons'' in the similar manner as in the $SU(2)$ case.\footnote{\,
Scattering $SL(2)$ (dyonic) giant magnon solutions can be constructed from the scattering $SU(2)$ (dyonic) giant magnon solutions $\xi_{i}(u_{1},u_{2};v_{1},v_{2})$ \cite{Spradlin:2006wk} by performing $(u_{1},u_{2})\mapsto (u_{1}+i\pi/2,u_{2}+i\pi/2)$\,.}

These ``shadow'' pictures remind us of the corresponding
finite-gap representations of both solutions, resulting from the
$SU(2)$ and $SL(2)$ spin-chain analyses. While in the $SU(2)$
case, a condensate cut, or a Bethe string, has finite length in
the imaginary direction of the complex spectral parameter plane,
for the $SL(2)$ case, they are given by two semi-infinite lines in
the same imaginary direction \cite{Minahan:2006bd}. This
complementary feature reflects the structural symmetry between the
BDS parts of S-matrices, $S_{SU(2)}=S_{SL(2)}^{-1}$\,.

These ``shadow'' pictures also show up in matrix model context
\cite{Berenstein:2005jq,Vazquez:2006hd,Hatsuda:2006ty,Berenstein:2007zf}. In a
reduced matrix quantum mechanics setup obtained from $\mathcal N =
4$ SYM on $\mathbb R\times S^{3}$\,, a ``string-bit'' connecting
eigenvalues of background matrices forming $\half$-BPS circular
droplet can be viewed as the shadow of the corresponding string.
For the $SU(2)$ sector, it is true even for the boundstate (bound
``string-bits'') case \cite{Hatsuda:2006ty}. It would be
interesting to investigate the $SL(2)$ case along similar lines of
thoughts.

\section{Useful Formulae\label{app:formula}}

This appendix provides some formulae useful for computation involving Jacobi elliptic functions and elliptic integrals.

\subsection{Elliptic Functions and Elliptic Integrals Near \bmt{k = 1}}\label{app:asymptotic}

The behavior of Jacobi elliptic functions around $k = 1$ is
discussed below.\footnote{\, We make the elliptic moduli explicit
in this section, and use the same conventions as
\cite{Okamura:2006zv}. } We follow the method of
\cite{Lin:1962:EJE}, where they computed asymptotics around $k =
0$\,.

\paragraph{$\bullet$ Jacobi sn, cn and dn functions.}

The Jacobi sn function obeys an equation
\begin{equation}
u = \int_0^{\sn(u,k)} \frac{dt}{\sqrt{1-t^2} \sqrt{1-k^2t^2}} \,.
\end{equation}
Differentiating both sides with respect to $k$\,, one finds
\begin{equation}
\frac{\partial \sn(u,k)}{\partial k} = - \cn (u,k) \dn (u,k) \int_{0}^{\sn(u,k)} \frac{k \ssp t^2 \, dt}{\sqrt{1-t^2} \pare{1-k^2t^2}^{3/2}} \,.
\end{equation}
Taking the limit $k \to 1$ and substituting $u=i\omega$\,, we obtain
\begin{equation}
\left. \frac{\partial \sn(u,k)}{\partial k} \right|_{k \to 1} = \frac{i \pare{\omega - \sin \omega \cos \omega}}{2 \cos^2 \omega} \,,
\end{equation}
which is the first term in the expansion of the Jacobi sn function
around $k=1$\,.

The asymptotics of the Jacobi cn and dn functions can be
determined by the relations
\begin{equation}
\sn^2 (u,k) + \cn^2 (u,k) = 1,\qquad \dn^2 (u,k) + k^2 \sn^2 (u,k) = 1\,.
\end{equation}

\paragraph{$\bullet$ Jacobi zeta function.}

The Jacobi zeta function behaves around $k=1$ as
\begin{equation}
Z_0 (u, k=e^{-r}) = \tanh u + \frac{z_2 (u)}{\ln r} + r \ssp z_1 (u) + \ldots.
\end{equation}
The functions $z_1 (u)$ and $z_2 (u)$ can be determined in the following way.
The third term, $z_1 (u)$\,, is calculated by the formula \cite{Byrd:1971:HEI}:
\begin{equation}
\mathop {\rm lim}_{k \to 1} \ \eK(k) \pare{Z_0 (u,k) - \tanh u} = - u\,,
\label{leading Zeta}
\end{equation}
while the second term, $z_2 (u)$\,, can be determined by the relations
\begin{equation}
\frac{\partial Z_0 (u,k)}{\partial u} = \dn^2 (u,k) - \frac{\eE (k)}{\eK (k)} \,,
\label{diff Zeta}
\end{equation}
and
\begin{equation}
Z_0 (u+v,k) - Z_0 (u,k) - Z_0 (v,k) = - k^2 \sn (u,k) \sn (v,k) \sn (u+v,k) \,.
\end{equation}

\paragraph{$\bullet$ Complete elliptic integrals.}

For actual use of the relations \eqref{leading Zeta} and \eqref{diff Zeta}, we need to know the asymptotics of complete ellitpic integrals.
They are given by
\begin{alignat}{2}
\eK (e^{-r}) &= - \frac12 \ssp \ln r + \frac32 \ssp \ln 2 - \frac14 \ssp r \ln r + o(r \ln^m r)\,, \ \ & & \\
\eE (e^{-r}) &= 1 - \frac12 \ssp r \ln r + o(r \ln^m r)\,, & &
\end{alignat}
with $m>1$\,. Changing the elliptic modulus from $k$ to
$e^{-r}$\,, the asymptotic behavior of elliptic functions around
$r=0$ are given by
\begin{alignat}{1}
\sn (\iom, e^{-r}) &= i \tan \omega - ir \, \frac{\omega - \sin \omega \cos \omega}{2 \cos^2 \omega} + O (r^2)\,,   \label{expand sn} \\[1mm]
\cn (\iom, e^{-r}) &= \frac{1}{\cos \omega} - r \frac{\omega \sin \omega - \sin^2 \omega \cos \omega}{2 \cos^2 \omega} + O (r^2)\,,  \label{expand cn} \\[1mm]
\dn (\iom, e^{-r}) &= \frac{1}{\cos \omega} - r \frac{\omega \sin \omega + \sin^2 \omega \cos \omega}{2 \cos^2 \omega} + O (r^2)\,,  \label{expand dn} \\[1mm]
Z_0 (\iom, e^{-r}) &= i \tan \omega - ir \, \frac{\omega + \sin \omega \cos \omega}{2 \cos^2 \omega} + \frac{2 \ssp i \ssp \omega}{\ln r} + O (r^2)\,.
\label{expand Zeta}
\end{alignat}

\subsection{Moduli transformations}\label{app:Ttransf}

We collect some formulae for $SL(2,\bb{Z})$ transformations acting
on elliptic functions.

Elliptic theta functions transform under the ${\sf T}$-transformation as
\begin{alignat}{2}
\vartheta _0 (z|\tau  + 1) &= \vartheta _3 (z|\tau )\,,&\qquad
\vartheta _1 (z|\tau  + 1) &= e^{\pi i/4} \, \vartheta _1 (z|\tau )\,, \\[1mm]
\vartheta _2 (z|\tau  + 1) &= e^{\pi i/4} \, \vartheta _2 (z|\tau )\,,&\qquad
\vartheta _3 (z|\tau  + 1) &= \vartheta _0 (z|\tau )\,,
\end{alignat}
and complete elliptic integrals with $q \ge 0$ transform as
\begin{equation}
\eK (q ) = k' \eK (k)\,,\qquad \eK '(q ) = k' \left( {\eK'(k) - i \eK (k)} \right)\,,\qquad \eE (q ) = \eE (k)/k'\,.
\end{equation}
Jacobian theta functions, defined by
\begin{equation}
\Theta _\nu  (z,k) \equiv \vartheta _\nu  \left( {\frac{z}{2 \eK(k)},\, \tau  = \frac{i \eK' (k)}{\eK (k)}} \right)\,, \quad
(\nu = 0,1,2,3)
\end{equation}
transform as
\begin{alignat}{2}
\Theta _0 (z|\tau  + 1) &= \Theta _3 ( z/k' |\tau )\,,&\qquad
\Theta _1 (z|\tau  + 1) &= e^{\pi i/4} \, \Theta _1 ( z/k' |\tau )\,, \\[1mm]
\Theta _2 (z|\tau  + 1) &= e^{\pi i/4} \, \Theta _2 ( z/k' |\tau )\,,&\qquad
\Theta _3 (z|\tau  + 1) &= \Theta _0 ( z/k' |\tau )\,,
\end{alignat}
and Jacobian zeta functions defined by $Z_\nu (z,k) \equiv \partial _z \ln \Theta _\nu (z,k)$ transform as
\begin{alignat}{2}
Z_0 (z|\tau  + 1) &= Z_3 (z/k' |\tau) / k' \,,&\qquad
Z_1 (z|\tau  + 1) &= Z_1 (z/k' |\tau) / k' \,, \\[1mm]
Z_2 (z|\tau  + 1) &= Z_2 (z/k' |\tau) / k' \,,&\qquad
Z_3 (z|\tau  + 1) &= Z_0 (z/k' |\tau) / k' \,.
\end{alignat}
Therefore, the ${\sf T}$-transformation acts on the elliptic
modulus $k$ as
\begin{alignat}{2}
q &\equiv \left( {\frac{{\Theta _2 (0|\tau  + 1)}}{{\Theta _3 (0|\tau  + 1)}}} \right)^2 = i \left( {\frac{{\Theta _2 (0|\tau)}}{{\Theta _0 (0|\tau)}}} \right)^2 & &= \frac{{ik}}{{k'}} \,, \\
q ^\prime &\equiv \left( {\frac{{\Theta _0 (0|\tau  + 1)}}{{\Theta _3 (0|\tau  + 1)}}} \right)^2 = \left( {\frac{{\Theta _3 (0|\tau)}}{{\Theta _0 (0|\tau)}}} \right)^2 & &= \frac{1}{{k'}} \,.
\end{alignat}
In terms of the modulus $q$ defined in (\ref{moduli kq}), the
Jacobian sn, cn and dn functions are written as
\begin{equation}
\sn(z,q ) = k' \frac{\sn (z/k',k)}{\dn (z/k',k)} \,, \quad
\cn(z,q ) = \frac{\cn (z/k',k)}{\dn (z/k',k)} \,, \quad
\dn(z,q ) = \frac{1}{\dn (z/k',k)} \,.
\end{equation}



\end{document}